\documentclass[10pt,tightenlines,eqsecnum,floats,aps,amsmath,amssymb,showpacs,nofootinbib,superscriptaddress,prd]{revtex4}

\usepackage{enumerate}
\usepackage{graphicx}
\usepackage{fontenc}
\usepackage{textcomp}
\usepackage{euscript,amssymb}
\usepackage{mathrsfs}

\usepackage{amsbsy}
\usepackage{epsfig}
\usepackage{color}

\usepackage{colordvi}

\newcommand{\scri}{{\mathscr I}}

\usepackage{amsfonts}
\usepackage{enumitem}
\usepackage{amsmath}

\usepackage{amssymb}

\usepackage{fancyhdr}

\usepackage{esint}
\usepackage[unicode=true, pdfusetitle,
bookmarks=true,bookmarksnumbered=false,bookmarksopen=false,
breaklinks=false,pdfborder={0 0 1},backref=false,colorlinks=false]
{hyperref}
\newcommand{\f}[2]{\frac{#1}{#2}}
\def\nn{\nonumber\\}

\usepackage{colordvi}

\begin{document}

\title{{\bf Null Fluid collapse in Rastall theory of gravity}}

\author{Amir Hadi Ziaie}
\email{ah.ziaie@maragheh.ac.ir} \affiliation{Research Institute for Astronomy and Astrophysics of Maragha (RIAAM), University of Maragheh, P.O. Box 55136-553, Maragheh, Iran}

\author{Yaser Tavakoli}
\email{yaser.tavakoli@guilan.ac.ir} 
\affiliation{Department of Physics,
University of Guilan, Namjoo Blv.,
41335-1914 Rasht, Iran}

\date{\today}

\begin{abstract}
A Vaidya spacetime is considered for gravitational collapse of a type II fluid in the context of Rastall theory of gravity. 
For a  linear equation of state ({\rm EoS}) for the fluid profiles,  the conditions under which the dynamical evolution of the collapse can give rise to the formation of a naked singularity  are examined. It is shown that depending on the model parameters, strong-curvature naked singularities would arise as  exact solutions to the Rastall's field equations. 
The allowed values of these parameters satisfy certain conditions on the physical reliability, nakedness and the curvature strength of the singularity.  It turns out that  Rastall gravity, in comparison to general relativity (GR), provides a wider class of physically reasonable spacetimes that admit both locally and globally naked singularities.
\end{abstract}

\date{\today}

\keywords{Naked singularities, black holes, modified theories of gravity, gravitational collapse}

\maketitle

\section{Introduction}

Since the mid-1970s when the singularity theorems were introduced by Hawking and Penrose~\cite{Hawking:1996jh}, the study of gravitational collapse and its final outcome has attracted a great interest in the theories of gravity and relativistic astrophysics (see also \cite{Hawking:1973uf}). 
As these theorems do not say much about the detailed features of the spacetime singularity, one would like to figure out that under what conditions the collapse can result in  formation of a black hole or a naked singularity. 
If the latter occurs as the collapse final state, it would be regarded as a counterexample to cosmic censorship conjecture (CCC)~\cite{Penrose:1969pc,wald1998black}. 
The CCC is a statement of the causal structure of spacetime during the gravitational collapse which indicates that the curvature singularities in an asymptotically flat spacetime will always be dressed by the event horizon of a black hole. Therefore, the singular regions cannot be causally connected to the external regions through an observer at infinity. 
This statement is usually referred to as the weak version of the CCC which implies the future predictability of the spacetime outside the black hole event horizons \cite{wald1984general}.
Nevertheless, the strong version of the CCC states that  singularities would not be visible to  observers unless they actually arrive there (see e.g.,~\cite{Clarke:1994,Wald:1997wa,earman1995bangs,Joshi:2013xoa} for reviews on CCC). Despite the fundamental role the CCC plays in the basic theory and applications of the black hole physics, possible violation of this hypothesis could be of significant importance.

Notwithstanding several attempts made by many researchers, neither a conclusive proof or disproof nor a precise and firm mathematical formulation of CCC has been presented up until now leaving this conjecture  as one of the most outstanding unresolved issues in GR. On the contrary, much efforts have been devoted in the past decades to find and develop exact spacetimes, as solutions to GR, which admit naked singularities as the collapse end product. This implies a situation for spacetime events where the apparent horizons fail to form as the collapse proceeds towards the singular region. Consequently, causal curves terminating in the past at the singularity have the chance to reach the faraway observers thus exposing super-dense regions of extreme gravity to the external universe~\cite{Joshi:1987wg,Joshi:2008zz}.
In this context, formation of naked singularities as collapse outcome has been studied within many different scenarios such as gravitational collapse of a dust cloud~\cite{Joshi:1993zg}, scalar fields' collapse~\cite{Maeda:1998hs,Giambo:2005se,Bhattacharya:2010tc,Tavakoli:2011iq,Nakonieczna_2015,Giamb__2009,Giambo:2008ya}, gravitational collapse in higher dimensions~\cite{Ghosh:2002mf,Goswami_2007,Giamb__2008,Yamada_2011,Dadhich_2013,Shimano_2014,Rahman:2018oso} etc. (for a review see \cite{Joshi:1987wg,Joshi:2008zz}).

Because of the generic singular feature of GR, it is believed that a more fundamental theory of gravity is needed to replace GR in the high energy regime. An  alternative candidate is the quantum theory of gravity which   governs the late time evolution of the gravitational collapse in the vicinity of the classical singularity.  It has been shown that  quantum effects resolve the classical singularity and  replace it by a quantum bounce~\cite{moore2006trends,Bojowald:2007ky,Tavakoli:2013tpa, Marto:2013soa} or by a   regular center eventuated out of an evaporation process \cite{Tavakoli:2013rna,Hayward:2005gi,Goswami:2005fu}. If the horizons form  before the collapse ends, such a quantum smeared region cannot be detected by an external observer \cite{Rovelli:2014cta}. Otherwise,  quantum gravity effects can be  carried out on the exterior region  providing  a feasible setting to study the high energy astrophysical phenomena.

One of the earliest examples of naked singularity was introduced by Papapetrou~\cite{Dadhich:1985zx} by considering  the collapse of a spherically symmetric  cloud  of radiation (i.e., a null fluid) represented by the Vaidya geometry equipped with a linear mass function \cite{Vaidya:1999zz}. 
This model  was extended later by generalizing the original Vaidya solution \cite{Wang:1998qx,Mkenyeleye:2014dwa}, namely, collapse of null strange quark fluid (NSQF)~\cite{Harko:2000ni,Ghosh_2003} and $N$-dimensional Vaidya spacetimes~\cite{Ghosh:2008zza},  in order to investigate the validity of the CCC thereon (for more references along this line see e.g.,~\cite{Husain:1995bf,Jhingan:2000xs,Lake:1991qrk,Dwivedi:1989pt,10.1143/PTP.72.63, Harko:2000ni,Ghosh_2003}). 
In the context of modified theories of gravity, violation of the CCC due to formation of a naked singularity was investigated, e.g., in the contexts of  $f({\cal R})$ gravity~\cite{Sharif:2010um,Ziaie:2011dh,Cembranos_2012}; Brans-Dicke theory~\cite{Hwang:2010aj,Ziaie:2010cz,Bedjaoui:2010nh}; Gauss-Bonnet gravity~\cite{Abbas:2017tvh,Narita:2009zza,Ghosh_2010,ZHOU_2011,Sharif:2013era}; Lovelock gravity~\cite{Ohashi_2011,Dadhich:2013bya,Zhou:2011vz}, $f({\cal R},T)$ gravity~\cite{G2019MPLA,Ahmed:2020xkh}; a modified gravity in the presence of spacetime torsion~\cite{Ziaie_2014,Ziaie_2014a,Luz_2018}; string theory~\cite{PhysRevD.70.104023,Gutperle_2004,Horowitz:2003yv}; asymptotically safe gravity~\cite{Bonanno:2017zen,platania2018asymptotically} and  improved Vaidya spacetimes based on renormalization-group~\cite{reuter2019quantum,Bonanno:2016dyv}.
Thereon, it was found that formation of naked singularities  depends on the various features of the theory.
Likewise,  the connection between CCC and weak gravity conjecture (WGC)~\cite{ArkaniHamed:2006dz} has been studied in~\cite{Horowitz:2019eum}. It was argued that if the WGC holds, the counterexamples to CCC in an asymptotically AdS case~\cite{Horowitz:2016ezu,Crisford:2017zpi} are not physical (see~\cite{Ong:2020xwv} for more details).
It is therefore of interest to investigate gravitational collapse in the framework of other theories of gravity with additional degrees of freedom, which are different from those provided by GR, and search for the possible existence of naked singularities.

Most of the modified theories of gravity  are described by a {\em divergence-free} energy-momentum tensor (EMT) which couples to the geometry in a minimal way~\cite{Lobo:2014ara,Capozziello:2010zz}. However, it is noteworthy that this property of the EMT, which leads to EMT conservation law, is not obeyed by the particle production process~\cite{Parker:1971pt}. 
Hence, it does not seem unreasonable to relax the condition on divergence-free EMT and search for a new gravitational theory and investigate its predictions. As it has been argued by Rastall~\cite{Rastall:1973nw}, we can examine the EMT conservation only locally or in a weak-field regime and this may not hold true in  curved spacetimes. Thus, in general, a non-trivial generalization of the EMT conservation  is   possible in principle. 
Based on Rastall's argument, a simple generalization of GR is to assume that the covariant divergence of EMT is directly proportional to the derivative of Ricci scalar, i.e., $\nabla_\mu T^\mu_{\,\,\,\, \nu}\propto\nabla_{\nu} {\mathcal R}$. As a result of this modification, there appears a non-minimal coupling between geometry and matter fields~\cite{Harko:2014gwa,Bertolami:2007gv,Rastall:1973nw}. In this sense, the ordinary EMT conservation law is written as $\nabla_\mu T^\mu_{\,\,\,\, \nu}=\chi\nabla_{\nu} {\mathcal R}$, where $\chi$ is introduced as the coupling parameter the limiting value $\chi\rightarrow0$ of which retrieves GR. 
An attracting feature of this theory is that the resulting field equations are much simpler than those of other curvature-matter theories and thus easier to inspect. 
Recently, Rastall gravity has attracted renewed interest in the literature due to its fascinating consequences in cosmological and astrophysical scenarios. For example, in Ref.~\cite{Heydarzade_2017} the charged and uncharged solutions of Kiselev-like black holes surrounded by a perfect fluid have been obtained. The Kiselev solutions refer to static spherically-symmetric exact solutions of Einstein field equation with the quintessential matter surrounding a black hole~\cite{Kiselev_2003}. Interesting and novel unexpected aspects of the cosmological models in the framework of Rastall theory have been investigated in~\cite{FABRIS_2012}. In~\cite{Moradpour:2017shy,AlRawaf:1995rs,Rahman1997,Arbab_2003,Batista:2010nq,Batista:2012hv}, it was shown that such a modification of GR is in a good agreement with observations as well as theoretical expectations. 
More interestingly,  Rastall  gravity, in comparison to  GR, provides somehow a better compatibility  with the observational data of the matter dominated era and the helium nucleosynthesis~\cite{AlRawaf:1994pn}. For studies of cosmological aspects of Rastall gravity, including its consistency with various cosmic eras, the reader can check the Refs.~\cite{Batista:2011nu,Majernik:2002gd}. Likewise, this theory provides a suitable setting to study the gravitational lensing effects~\cite{AbdelRahman:2001pb}. 

Nevertheless, in a recent paper~\cite{Visser:2017gpz} Visser argued that Rastal gravity is equivalent to GR. In his sight, Rastall gravity is simply a trivial rearrangement of the matter sector in GR while the geometrical part of the field equations for both theories are identical.
Moreover, having considered the standard variational principle in order to find a physically conserved EMT, the construction of Rastall EMT from a {\it Rastall action} is questionable. Then, to get the Rastall EMT through the standard variational principle one needs to introduce some non-dynamical background field~\cite{Gratus_2012}. Even if this is feasible, the Einstein-Hilbert action would be unaffected by such construction whatsoever and the gravity sector remains the same as that of the standard GR. This issue has been dealt with by Darabi et al.~\cite{Darabi_2018} in an objective way. Indeed, according to Visser~\cite{Visser:2017gpz}, the Rastall\rq{}s definition of	EMT is incorrect because the ordinary EMT differs from the generated effective one. However, Darabi et al. asserted that this argument cannot be true, as otherwise one could generalize this recipe to all other modified  theories of gravity concluding that these theories  can be  simply generated by special rearrangements of the Einstein's field equation, which, of course, is not true. They further argued that the EMT considered by Visser is not physically relevant while the definition of the EMT considered in the Rastall theory is the ordinary one.
To support their claim, Darabi et al. provided an example of a well-established $f({\mathcal R})$ gravity by generalizing the Visser's approach. After some rearrangements, they concluded that $f({\mathcal R})$ theory of gravity can also be written effectively  correspondent to GR but  not essentially equivalent to it.

From cosmological perspective, the differences between Rastall gravity and GR have been also reported in~\cite{PhysRevD.96.123504} and the earlier work~\cite{Smalley_1983} which allude to the non-equivalence of these two theories. 
Moreover, the effects of the  Rastall's coupling parameter on modeling the static and spherically symmetric distributions of a perfect fluid  have been surveyed in~\cite{Hansraj_2019}, where the properties of a well-known stellar model proposed by Tolman~\cite{PhysRev.55.364} was  investigated. Therein, it was shown  that, in most of the case studies, Rastall theory remains well-consistent with the basic requirements for physical reliability of the model while the GR theory exhibits defective behavior. The findings of this study and the similar works in Refs.~\cite{PhysRevD.92.044020,Abbas_2018,ABBAS20201} show that the Rastall's coupling parameter can act as a mathematical tool to compensate for the shortcomings of the standard GR. However, despite all comments for and against Rastall gravity, the question of its non-equivalence to GR is still under debate and more investigations are needed. 

Recently, gravitational collapse of a spherically symmetric homogeneous perfect fluid in Rastall gravity has been studied in~\cite{Ziaie_2019}. Therein, by considering a FLRW metric for the interior spacetime of the collapsing body equipped with a linear EoS for fluid profiles, it was found that for  certain values of EoS parameter  naked singularities can  arise as collapse end state whereas in GR the same matter profiles lead to the  black hole formation. 
Moreover,   it was shown that, in contrast to GR, an exterior Schwarzschild black hole can form as the collapse end state with a non-vanishing fluid pressure.
Motivated by these results and also the arguments provided by the  paragraphs above, our aim in the present work is to study gravitational collapse of a null fluid within the framework of Rastall gravity and examine the conditions under which the model parameters allow the formation of a naked singularity. The class of spacetimes that are natural candidates for modeling of such type of matter fields are the generalized Vaidya spacetimes. The matter field in these spacetimes have two components: a general type-I matter field and a type-II matter field describing the zero rest mass fields (null radiation). Our analyses will show that, in Rastall gravity, there can be found a wider class of physically reasonable spacetimes admitting naked singularities and hence more counterexamples to the CCC in comparison to the case of GR. 

Our paper is then organized as follows:
In Sec.~\ref{RastalFieldEq} we will present the field equations for the Vaidya spacetime in the presence of a null fluid within the context of Rastall  gravity.
In Sec.~\ref{Outcome-collapse} we will establish the physically reasonable conditions for the dynamical  evolution of the collapse within our model. Our main concern in this section  will be to investigate the required circumstances for the formation of  naked singularities as collapse outcome. 
Finally, in Sec.~\ref{Conclusion} we will provide the conclusions and discussions of our work.

\section{Gravitational collapse in Rastall gravity}
\label{RastalFieldEq}

The field equations of Rastall theory are given as~\cite{Rastall:1973nw,AlRawaf:1995rs}
\begin{eqnarray}
G_{\mu\nu} +\gamma g_{\mu\nu}{\mathcal R}=\kappa T_{\mu\nu},
\label{RastallFES}
\end{eqnarray}
where $\gamma=\kappa\chi$  is the Rastall dimensionless parameter and $\kappa$ is the Rastall gravitational coupling constant. 
By introducing an effective EMT, $T^{\rm eff}_{\mu\nu}$:
\begin{equation}
T^{\rm eff}_{\mu\nu} := T_{\mu\nu}-\f{\gamma T}{4\gamma-1}g_{\mu\nu},
\label{T-eff}
\end{equation}
the   equation above can be rewritten in an equivalent form as
\begin{eqnarray}\label{FESEquiv}
G_{\mu\nu} =\kappa T^{\rm eff}_{\mu\nu},~~~~~~~~~\kappa=\f{4\gamma-1}{6\gamma-1}8\pi G.
\end{eqnarray}
where $G$ is the universal gravitational constant and the last expression comes from the Newtonian limit of the Rastall theory~\cite{Rastall:1973nw,Capone_2010,Moradpour:2016fur}

For a spherically symmetric collapse of a null fluid we consider the Vaidya metric in Eddington-Finkelstein coordinates as
\begin{eqnarray}\label{vaidyametr}
ds^2 =-\left(1-\f{2m(r,v)}{r}\right)dv^2+2\epsilon drdv+r^2d\Omega^2, \quad \quad
\end{eqnarray}
where $m(r,v)$ is the mass function related to the gravitational energy confined within the radius $r$ and $d\Omega^2=d\theta^2+\sin^2\theta d\phi^2$ is the standard metric on a unit two-sphere. For $\epsilon=+1$, the null coordinate $v$ is the Eddington advanced time for which the $r$ coordinate is decreasing towards the future along a ray with $v=const.$  and represents a congruence of ingoing light rays.  Likewise, the case $\epsilon=-1$ represents a congruence of outgoing light rays. 

The EMT of the collapsing body is assumed to be given by a two-fluid system as
\begin{eqnarray}\label{EMTnm}
T_{\mu\nu}=T^{(\rm n)}_{\mu\nu}+T^{(\rm m)}_{\mu\nu},
\label{typeII}
\end{eqnarray}
where
\begin{eqnarray}
T^{(\rm n)}_{\mu\nu}\ &=&\ \sigma n_\mu n_\nu,
\label{TMN}\\
T^{(\rm m)}_{\mu\nu}\ &=&\ (\rho+p)(n_\mu \ell_\nu+n_\nu \ell_\mu)+pg_{\mu\nu}.
\label{TMN1}
\end{eqnarray}
The first relation, $T^{(\rm n)}_{\mu\nu}$, denotes the EMT of a null radiation which corresponds to the component of the matter field moving along the null hypersurfaces $v = const.$, while the second relation, $T^{(\rm m)}_{\mu\nu}$,
is the EMT of an ordinary matter. By considering a congruence of ingoing light rays (i.e.,  by setting $\epsilon=+1$), we define the vectors $n_\mu$ and $\ell_\nu$ as two null vector fields
\begin{eqnarray}
n_\mu=\delta_\mu^{\,0}, \quad \quad \ell_\nu=\f{1}{2}\left[1-\f{2m(r,v)}{r}\right]\delta_\nu^{\,0}+\delta_\nu^{\,1}, \quad \quad~~~~~n_\lambda n^\lambda=\ell_\lambda\ell^\lambda = 0, \quad \quad n_\lambda \ell^\lambda=-1.
\label{ellk}
\end{eqnarray}

Let us assume that the {\rm EoS} of the two-fluid system is of the form $p=w\rho$. Setting this in the field equation~(\ref{FESEquiv}) we arrive at a second order differential equation for $m(r,v)$. By solving this equation  we get the following relations for the mass function:
\begin{eqnarray}
m \left( r,v \right) = \left\{
\begin{tabular}{cc}
$\alpha F_1(v)r^\beta+M(v)$, & \,  if   $\gamma\neq\f{2w-1}{2(1+w)}$, \\
\\
$F_1(v)\ln(r)+M(v)$, & \,  if $\gamma=\f{2w-1}{2(1+w)}$,    \\
\end{tabular} \right. \quad   
\label{solmassf}
\end{eqnarray}
where, $F_1(v)$ and $M(v)$ are two arbitrary functions of integration and $\beta$ and $\alpha$ are defined by
\begin{eqnarray}
\beta(\gamma,w) =-1-\f{2(1-w)}{2\gamma(1+w)-1} =: \f{1}{\alpha(\gamma, w)}\, . \quad \
\label{beta}
\end{eqnarray}
In case $p=\rho=0$, the {\rm EMT} (\ref{EMTnm}) reduces to that of a pure radiation fluid. For this particular case we have $F_1(v)=0$ and the original Vaidya solution with the mass $m(r,v)=M(v)$ is retrieved~\cite{Wang:1998qx}. Moreover, for $p=\rho$ and $\gamma\neq1/4$, we have $\beta=-1$ and the solution (\ref{solmassf}) reduces to the Bonnor-Vaidya charged solution~\cite{Bonnor:1970zz}. We also note that in the limit $\gamma\rightarrow0$ this solution reduces to 
\begin{eqnarray}
m \left( r,v \right) = \left\{
\begin{tabular}{cc}
$M(v)-\frac{F_1(v)}{(2w-1)r^{2w-1}}$, & \,  if   $w\neq\frac{1}{2}$, \\
\\
$M(v)+F_1(v)\ln(r)$, & \,  if $w=\frac{1}{2}$,    \\
\end{tabular} \right. \quad   
\label{solmassf1}
\end{eqnarray}
which is general relativistic limit of the theory~\cite{Husain:1995bf}. By projecting the effective {\rm EMT} into the orthonormal basis:
\begin{eqnarray}\label{projecttetra}
{\sf e}_{(a)\mu}=\begin{pmatrix} 
\f{\sqrt{2}}{2}\left(\f{3}{2}-\f{m}{r}\right) & -\f{1}{\sqrt{2}} & 0 & 0\\
\f{\sqrt{2}}{2}\left(\f{3}{2}+\f{m}{r}\right) & \f{1}{\sqrt{2}} & 0 &0\\
0 & 0& r &0\\ 
0 & 0& 0 &r\sin\theta
\end{pmatrix}, \quad \quad 
\end{eqnarray}
we find that $T^{{\rm eff}(a)(b)}\equiv{\sf e}^{(a)\mu}{\sf e}^{(b)\nu}T^{\rm eff}_{\mu\nu}$ takes the form
\begin{eqnarray}
T^{{\rm eff}(a)(b)}=\begin{pmatrix} 
\frac{\tilde{\sigma}}{2}+\tilde{\rho} & \frac{\tilde{\sigma}}{2} & 0 & 0\\
\frac{\tilde{\sigma}}{2} & \frac{\tilde{\sigma}}{2}-\tilde{\rho} & 0 & 0\\
0 & 0& \tilde{p} &0\\ 
0 & 0& 0 &\tilde{p}
\end{pmatrix}, \quad \quad
\label{projectedemt}
\end{eqnarray}
which is the {\rm EMT} of a type II fluid as defined in Ref.~\cite{Hawking:1973uf}. 

As we discussed earlier, the CCC is a fundamental cornerstone of black hole physics.   As  noted by Penrose~\cite{Penrose:1969pc,wald1998black}, the key idea behind the CCC is the physically reasonableness, that is, a complete gravitational collapse of physically reasonable matter fields always leads to the formation of a black hole rather than a naked singularity. It is therefore generally expected that naked singularities must be developed under reasonable physical conditions on matter sources, otherwise, these singularities cannot be taken as serious counterexamples to CCC~\cite{JOSHI_2011,Joshi:1987wg,Joshi:2008zz}.
Despite the arbitrariness of the functions $F_1(v)$ and $M(v)$ in our  solution (\ref{solmassf1}), they should be chosen carefully so that the mass function $m(r,v)$ provides a physically reasonable EMT (\ref{projectedemt}) satisfying certain conditions on positivity of matter energy density and its dominance over the pressure. These requirements which are conventionally referred to as energy conditions are summarized as {\em weak}, {\em null}, {\em strong}  and  {\em dominant} energy conditions. 
	
The weak energy condition (WEC) asserts that for all timelike vector fields $v^\mu$, the EMT has to satisfy $T_{\mu\nu}v^\mu v^\nu\geq0$. This  implies that the energy density as measured by any observer is locally non-negative~\cite{Hawking:1973uf,poisson_2004}. For the EMT (\ref{projectedemt}) the WEC gives rise to the following constraints~\cite{Hawking:1973uf,Wang_1999}
\begin{equation}\label{wec}
\tilde{\sigma}\geq0,\, \, \,~~~\tilde{\rho}\geq0,\, \, \,~~~\tilde{p}\geq0.
\end{equation}
The WEC implies the null energy condition (NEC) which states that $T_{\mu\nu}k^\mu k^\nu\geq0$ for all null vector fields $k^\mu$~\cite{poisson_2004}. Likewise, the dominant energy condition (DEC) requires that for any future-directed timelike vector field $v^\mu$, $T_{\mu\nu}v^\mu v^\nu\geq0$ and the quantity $T_{\mu\nu}v^{\mu}$, which is the matter's momentum density  measured by an observer with the four-vector velocity $v^\mu$, is neither past-directed nor spacelike. Physically, this condition implies that the local energy flow cannot travel faster than light. In terms of the effective EMT  (\ref{projectedemt}), the DEC leads to the following inequalities~\cite{Hawking:1973uf,Wang_1999}:
\begin{equation}\label{dec}
\tilde{\sigma}\geq0,\, \, \,~~~\tilde{\rho}\geq \tilde{p}\geq0.
\end{equation}
Finally, the strong energy condition (SEC) states that $\left(T_{\mu\nu}-\frac{1}{2}Tg_{\mu\nu}\right)v^\mu v^\nu\geq0$ for any time like vector field $v^\mu$. From physical point of view, this condition is interpreted as the attractive nature of gravity. We also note that both WEC and SEC are identical for a Type II fluid.  
For a physically reliable collapse scenario here, we thus require that the mass function $m(r,v)$ is suitably chosen so that the energy conditions (ECs) (\ref{wec}) and (\ref{dec}) are respected. This would constrain the statements for $F_1(v)$ and $M(v)$ as well as the parameters $\gamma$ and $w$.

Using the mass function (\ref{solmassf}) we can rewrite the effective profiles as 
\begin{eqnarray}
\tilde{\sigma}(r,v) &=& \f{2(6\gamma-1)}{(4\gamma-1)r^2}\left(\dot{M}(v)+\alpha r^\beta \dot{F}_1(v)\right),\label{effprofs0}\\
\tilde{\rho}(r,v) &=& \f{2(6\gamma-1)}{4\gamma-1}F_1(v)r^{\beta-3},  \quad \label{effprofs1}\\
\tilde{p}(r,v) &=& \f{(1-\beta)(6\gamma-1)}{4\gamma-1}F_1(v)r^{\beta-3}, \quad \quad \quad \quad \label{effprofs}
\end{eqnarray}
where $\gamma\neq\f{1}{6}$ and $\gamma\neq\f{1}{4}$.
Moreover, we have set the units where $8\pi G=1$.
As we shall consider in Subsec. \ref{localnaked},  the mass function (\ref{solmassf}) is a growing function due to the inward flow of null fluid. We therefore take the functions $F_1(v)$ and $M(v)$ to be positive and increment in $v$. Then, to satisfy the ECs, the coefficients of these functions and their derivatives are required to be positive. Now, following the  arguments above together with the conditions  (\ref{wec}) and (\ref{dec})  the pair $\{\gamma , w\} $ are constrained to  the following sets of ranges: 
\begin{equation}
\{\gamma , w\} \in \bigvee\limits_{i=1}^4 S_i\, ,
\label{ex1012-1}
\end{equation}
where,  $S_i$'s are given by
\begin{eqnarray}
S_1&=&  \Big\{\gamma<\f{1}{6}\quad {\rm and}\quad  \mathrm{w}_\gamma\leq w<\tilde{\mathrm{w}}_\gamma\Big\},   \nonumber \\
S_2&=& \Big\{\f{1}{4}<\gamma<\f{1}{2}\quad {\rm and}\quad \tilde{\mathrm{w}}_\gamma<w\leq \mathrm{w}_\gamma\Big\}, \nonumber\\
S_3&=& \Big\{\f{1}{2}\leq \gamma<1\quad {\rm and}\quad  w> \tilde{\mathrm{w}}_\gamma\Big\}, \nonumber\\
S_4&=& \Big\{\f{1}{2}<\gamma\leq 1\quad {\rm and}\quad w\leq \mathrm{w}_\gamma\Big\},  
\label{ex1012}
\end{eqnarray}
where,  for any given  value of $\gamma$, the parameters $\mathrm{w}_\gamma$ and $\tilde{\mathrm{w}}_\gamma$ are constants and are defined by
\begin{eqnarray}
\mathrm{w}_\gamma:= -\f{2\gamma}{2\gamma-1}\, , \,  \quad \,   \tilde{\mathrm{w}}_\gamma:= -\f{1+2\gamma}{2(\gamma-1)}\, . \quad
\end{eqnarray}

\section{The status of final singularity}
\label{Outcome-collapse}

The formation of  naked singularity or  black hole  and the conditions under which any of these two outcomes would occur are of great significance in the study of gravitational collapse. 
The occurrence of these outcomes  can be examined via investigating the behavior of  outgoing light rays in the neighborhood of the singular region. 
In particular, if there exist families of null geodesics terminating at the singularity in the past,  whose tangent vectors are  positive-definite, the singularity can be revealed to the exterior observer  by meeting such curves. Otherwise, the singularity would be covered through a black hole horizon. 
Nevertheless, existence of such  rays do not guarantee that the singularity will be certainly naked. Indeed, to reach the exterior universe, it should be ensured that these outgoing  rays will not  be trapped within their journey due to formation of apparent horizons. This requires a  careful analysis of  the evolution of trapped surfaces  prior to  formation of the singularity.

Motivated from  paragraph above, our aim in this section will be studying the conditions under which the collapse scenario of a two-fluids system in Rastall gravity leads to the formation of a naked singularity. To be more concrete, the required properties for our collapsing model to be qualified in the sequel are:
\begin{enumerate}[label=(\roman*)]
	\item The effective EMT of the two-fluid system should satisfy the energy conditions so that its associated parameters $\{\gamma , w\}$   satisfy the ranges (\ref{ex1012-1}); \label{I}
	\item there should exist at least one  non-negative, real root for the  null geodesic equation (\ref{rooteq}): in fact, such roots are  tangents to null geodesics originating from the singularity (cf. Subsec. \ref{localnaked}); \label{II}
	\item the tangent vectors to the radial null geodesics, $(dv/dr)_{r,v=0}$, should be smaller than the slope of trapping surfaces, $(dv/dr)_{\rm AH}$, (cf. Eq.~(\ref{slopeah})) in the vicinity of the singularity. This ensures that the outgoing null geodesics will not be trapped due to the formation of apparent horizons.\label{III}
	
	\item  and finally, the emergent singularity at the collapse end state should be gravitationally strong so that the extension of spacetime through it cannot  be possible. \label{IV}
\end{enumerate}

Once the  conditions above are fulfilled, we will provide in Subsec. \ref{Results}, a numerical analysis of the physically relevant solutions for the herein collapse model and study the situations where a naked singularity can form as the collapse end product.

\subsection{Existence of outgoing radial null geodesics}
\label{localnaked}

The situation being considered here is that of a Vaidya spacetime (\ref{vaidyametr})  with a radially injected flow of radiation in it initiating from an empty region of Minkowski spacetime~\cite{Joshi:1987wg,Joshi:2008zz}. The radiation could have emerged from a central singularity at $r=0$, $v=0$ in the past with a growing Vaidya mass  $m(r,v)$. Hence, the Vaidya mass $m(r, v)$  is an arbitrary non-negative increment function of the radial coordinate and the advanced time. The first step in manifestation of the singular region is the  existence of outgoing non-spacelike geodesics originating from the singularity. In~\cite{Nolan_2002,Maeda_2006}, it is shown that if a future-directed radial null geodesic does not emanate from the singularity, then a future-directed causal (excluding radial null) geodesic does not either. Hence, for our purpose here, it suffices to consider only radially outgoing future-directed null geodesics originating at the singularity. Let us define $\zeta^\mu=dx^\mu/d\eta$ where $x^\sigma=(v, r, \theta, \phi)$, as the tangent to null geodesics with $\eta$ being an affine parameter. The null condition $\zeta^\mu\zeta_\mu=0$ for radial geodesics $[\zeta^\mu]=(\zeta^v, \zeta^r,0,0)$ gives
\begin{eqnarray}
\left(1-\f{2m}{r}\right)\left(\zeta^v\right)^2-2\zeta^v\zeta^r=0,
\label{nullcond}
\end{eqnarray}
whereby defining $\zeta^v=Q(r,v)/r$, with $Q(r,v)$ being an arbitrary function~\cite{Joshi:1987wg,Dwivedi:1989pt}, we get the following relation for $\zeta^r$ as
\begin{eqnarray}\
\zeta^r=\left(1-\f{2m}{r}\right)\f{Q}{2r}.
\label{UrQ}
\end{eqnarray}
It is now useful to define tangent to the radial null geodesics given by the following relation~\cite{Dwivedi:1989pt,Joshi:1987wg,Mkenyeleye:2014dwa}
\begin{eqnarray}
Z\equiv\frac{\zeta^v}{\zeta^r}= \frac{dv}{dr} = \frac{2r}{r-2m}.\, 
\end{eqnarray}
Indeed, for any family of non-spacelike curves meeting the singularity,
the tangents to the curves are definite \cite{Mkenyeleye:2014dwa}. This means that the parameter $Z$ given in the limit  $r=0$, $v=0$:
\begin{eqnarray}\label{Z0-1}
Z_0 &:=& \lim_{v,r\rightarrow0}Z = \left(\f{dv}{dr}\right)_{v,r=0}\, ,
\end{eqnarray}
is well-defined at the singularity. For the given  metric (\ref{vaidyametr}), Eq.~(\ref{Z0-1}) becomes
\begin{eqnarray}
Z_0 &=& \f{2}{(1-2m_0^\prime)-2\dot{m}_0Z_0},
\label{Z0-2}
\end{eqnarray}
where we have defined 
\begin{eqnarray}
\dot{m}_0 := \left(\f{\partial m}{\partial v}\right)_{v,r=0}\quad {\rm and} \quad {m}^\prime_0 := \left(\f{\partial m}{\partial r}\right)_{v,r=0} . \quad 
\label{mpm0}
\end{eqnarray} 
The status of final singularity is determined by the characteristic (limiting) parameter  $Z$  on the singular geodesics. This 
is given  by solving the quadratic  equation (\ref{Z0-2}) for $Z_0$: 
\begin{eqnarray}\label{Z0}
2\dot{m}_0Z_0^2 -(1-2m_0^\prime)Z_0 +2 = 0.
\end{eqnarray}
We thus obtain 
\begin{eqnarray}
Z_0=\f{1-2m_0^\prime\pm\left[(1-2m_0^\prime)^2-16\dot{m}_0\right]^{\f{1}{2}}}{4\dot{m}_0}.
\label{Z0solve}
\end{eqnarray}
Now, to determine the limiting  parameter $Z_0$ we should specify the mass function $m(r,v)$ in equation above.

A suitable choice for the mass function $m(r, v)$ and its partial derivatives ensures the existence of the well-defined solutions for the  tangents $Z_0$ to the null geodesics in the vicinity of the singularity. To be more precise, positive-definiteness of the $Z_0$ in Eq.~(\ref{Z0solve})  implies that the partial derivatives of the mass function should exist and be continuous on the entire spacetime of the collapse. Moreover, they should hold the conditions $(1-2m_0^\prime)^2-16\dot{m}_0\geq0$ and $\dot{m}_0>0$ (provided by the requirement that $\dot{m}\neq0$ and the  weak energy condition $\sigma(r,v)>0$ is satisfied) at the central singularity. Having the solutions for $Z_0$ with the  properties above 
guarantees the  existence of families of future directed non-spacelike trajectories that {\em can} reach  faraway observers in spacetime.

In order to determine the tangents $Z_0$ we then proceed with computing the mass function $m(r, v)$  in the limit when the singularity is reached. In particular, we follow Ref.~\cite{Joshi:1987wg} and consider an influx of null fluid  collapses to the singularity, where the first shell arrives at $r=0$ at time $v=0$ while the last shell arrives at $v=v_0$. We further consider a situation where  the radial null fluid starts its evolution, for $v<0$, from an initially empty region of the Minkowski spacetime where $m(r,v)=0$. 
For $v>v_0$ we would have a Schwarzschild spacetime with a constant mass $m(r,v)=M_0$.
Then, the suitable expressions for the arbitrary functions $F_1(v)$ and $M(v)$ can be given by
\begin{eqnarray}
F_1(v) := \left\{ \begin{array}{lll}
0\, , & \quad \mbox{if \, $v < 0$}\, ;\\
\xi v^{1-\beta}, & \quad \mbox{if \, $0\leq v \leq v_0$}\, ;\\
0\, ,& \quad \mbox{if \, $v > v_0$}\, ,\end{array} \right. 
\label{F1v}
\end{eqnarray}
and
\begin{eqnarray}
M(v) := \left\{ \begin{array}{lll}
0\, , & \quad\mbox{if \, $v < 0$}\, ;\\
\lambda v\, , & \quad \mbox{if \, $0\leq v \leq v_0$}\, ;\\
M_0\, ,& \quad \mbox{if \, $v > v_0$}\, ,\end{array} \right.
\label{F2v} 
\end{eqnarray}
where, $\lambda$ and $\xi$ are some constants.   
Using this choice in Eq.~(\ref{solmassf})  gives the mass function
as
\begin{eqnarray}\
m(r,v) = \left\{ \begin{array}{lll}
0 \, , & \quad \mbox{if \, $v < 0$}\, ;\\
\frac{\xi}{\beta} v^{1-\beta}r^\beta +\lambda v\, , & \quad \mbox{if \, $0\leq v \leq v_0$}\, ;\\
M_0\, ,& \quad \mbox{if \, $v > v_0$}\, .\end{array} \right. \quad \quad 
\label{solmassf-2}
\end{eqnarray}
The limiting values for the partial derivatives of the mass function (\ref{solmassf-2}) in the vicinity of the singularity now read
\begin{eqnarray}
m^\prime_0=\xi Z_0^{1-\beta} \quad {\rm and} \quad \dot{m}_0=\lambda+\frac{1-\beta}{\beta}\xi Z_0^{-\beta}. \quad \quad 
\label{mdotpr0}
\end{eqnarray}
By replacing $m^\prime_0$ and $\dot{m}_0$ from equation above into the  Eq.~(\ref{Z0solve}) we get the root equation as
\begin{eqnarray}
\f{2\xi}{\beta}Z_0^{n}+2\lambda Z_0^2-Z_0+2=0, 
\label{rooteq} 
\end{eqnarray}
where 
\begin{eqnarray}
n(\gamma,w) := 2-\beta(\gamma,w)\, .
\label{n-beta}
\end{eqnarray}
Eq.~(\ref{rooteq}) is indeed an algebraic equation whose solutions represent 
the behaviour of  the outgoing null rays responsible for revelation of the central singularity to the exterior universe. To find the desired solutions to this equation we  first need to determine the exponent $n(\gamma,w)$ due to the physically reasonable ranges of parameter $\{\gamma, w\}$.
This follows from qualification of the remaining conditions (i.e., items \ref{III} and \ref{IV}) listed at the beginning of this section.
\par
So far we have established, due to Eq.~(\ref{rooteq}), a situation for the existence of outgoing null rays, through positive-definite tangent $Z_0$, as a functor for revelation of the spacetime singularity. However, this does not guarantee the nakedness of the final singularity. Indeed, if  apparent horizons form early enough prior to the singularity formation, the outgoing rays will be trapped and the singularity would be covered by a black hole horizon. To avoid such situations, our task would be now to set up an evolution equation for the trapping horizons and examine circumstances for the null geodesics to stand outside the trapped region. To this aim we first write the expansion parameters along the null vector fields $n_\nu, \ell_\mu$ as~\cite{York:1984,Faraoni:2018xwo}
\begin{eqnarray}
\Theta_\ell(r,v)=\f{r-2m(r,v)}{r^2},\quad \quad \Theta_n(r,v)=-\f{2}{r}. \quad \quad 
\label{expanelln}
\end{eqnarray}
For  spheres  in $r<2m(r,v)$ region,  we  get  both  expansions  to  be  negative. Such  spheres  are  known  as trapped surfaces whose  union form a trapped region. Thus,  apparent horizon is the boundary of the trapped region and is defined by the conditions
\begin{eqnarray}
\Theta_\ell=0\quad {\rm and} \quad  \Theta_n<0.
\label{aphcond}
\end{eqnarray}
From this,  the equation for the apparent horizon  can be written in the form
\begin{eqnarray}\label{AHEQ}
2m(r,v)=r.
\label{aphoreq}
\end{eqnarray}
Now, we can calculate the slope of apparent horizon in the limit where  the singularity is approached. By differentiating Eq.~(\ref{aphoreq}) we obtain $2dm(r,v)=dr$, which can be written as
\begin{eqnarray}\label{diffapphorcurve}
\dot{m}(r,v)\, \left(\f{dv}{dr}\right)_{{\rm AH}}+m^\prime(r,v)=\frac{1}{2}\, .
\end{eqnarray}
Here, the term $(dv/dr)_{\rm AH}$ represents the tangent to the trapping surfaces.
Therefore, by solving the above equation for the slope of  apparent horizon, $X_0\equiv(dv/dr)_{\rm AH}$, in the limit where $v, r\rightarrow0$ (i.e., as the singularity is approached), we  get 
\begin{eqnarray}\label{slopeah}
X_0\ =\  \f{1-2\xi Z_0^{1-\beta}}{2\lambda+\f{2\xi}{\beta}(1-\beta)Z_0^{-\beta}}\, . \quad 
\end{eqnarray}
In derivation of equation above, we have used Eq.~(\ref{mdotpr0}) for the values of $\dot{m}_0$ and $m^\prime_0$ in the vicinity of the singularity.
Therefore, once a positive  solution of $Z_0$ is given due to Eq.~(\ref{rooteq}), the slope of the apparent horizon, $X_0$, would be determined. If  the tangents to the outgoing null geodesics are smaller than the slope of the apparent horizon in the vicinity of the central singularity, i.e. $Z_0<X_0$, then the outgoing rays will lie outside the trapped region and the final singularity would  be naked. Depending on the qualified values of parameters $\{\gamma, w, \xi, \lambda\}$, in Subsec. \ref{Results}, we will provide analyses to examine the existence of our favorite solutions satisfying the condition $Z_0<X_0$ for the nakedness of the spacetime singularity. 

\par
Next, we proceed to examine the curvature strength of the spacetime singularity. According to Tippler \cite{Tipler:1977zza}, an important test of the physical significance of a spacetime singularity is its curvature strength. Indeed, if singularity is gravitationally weak, then, extension of the spacetime through it may be possible. On the contrary, when a strong curvature singularity forms, the gravitational tidal forces associated with it are so strong that any object trying to cross it gets destroyed. Therefore, as argued by~\cite{Ori:1991zz}, the extension of spacetime becomes meaningless for such a strong curvature singularity for which all the objects terminating at it shrink to zero size. In order to estimate the curvature strength of the singularity we consider a congruence of null geodesics parameterized by the affine parameter $\eta$ that terminate at the singularity. Then, the singularity would be gravitationally strong if the following condition holds~\cite{Clarke:1985}:
\begin{eqnarray}
\psi &=& \lim_{\substack{\eta\rightarrow0}}\, \eta^2R_{\mu\nu}\zeta^\mu\zeta^\nu>0.
\label{curvstr}
\end{eqnarray}
To compute the  quantity above, we consider the required components for the Ricci tensor as
\begin{eqnarray}\label{Riccitens} 
R_{vv}\ &=&\ (2m-r)\f{m^{\prime\prime}}{r^2}+\f{2\dot{m}}{r^2}\, ,\nn
R_{vr}\ &=&\ \f{m^{\prime\prime}}{r}\, .
\end{eqnarray} 
Then, using this together with the tangent vector field $\zeta^\mu$ through Eq.~(\ref{curvstr}) we get
\begin{eqnarray}
\psi = \lim_{\substack{\eta\rightarrow0}} \, 2\dot{m}(r,v)\left(\f{Q\eta}{r^2}\right)^2 = 2\dot{m}(r,v)\lim_{\substack{\eta\rightarrow0}}\left(\f{dv}{d\eta}\right)^2\f{\eta^2}{r^2}\, .
\label{Psi0}
\end{eqnarray}
Using now the l'H\^{o}pital's rule in equation above, we get the following expression in the limit of approach to the singularity as
\begin{eqnarray}
\psi &=& \dot{m}_0 \lim_{\substack{\eta\rightarrow0}}\left(\f{dv}{dr}\right)^2\, =\,  2\dot{m}_0Z_0^2\, .
\label{Psi-1}
\end{eqnarray}
Now, by setting $\dot{m}_0$ from Eq.~(\ref{mdotpr0}) into equation above, we obtain
\begin{eqnarray}
\psi &=& 2\lambda Z_0^2+\f{2\xi}{\beta}(1-\beta)Z_0^{2-\beta}.
\label{Psi}
\end{eqnarray}
It therefore follows that depending on the model parameters, $\psi$ can be positive and the strong curvature condition (\ref{curvstr}) is satisfied.  We shall examine the behavior of this parameter in the next subsection.

\subsection{Numerical results and fate of the singularity}
\label{Results}

Once the conditions \ref{I}--\ref{IV} are met, the herein model for null fluid collapse in Rastall gravity is qualified whose central singularity can be visible to a distant observer.

\begin{figure}
	\begin{center}
		\includegraphics[scale=0.24]{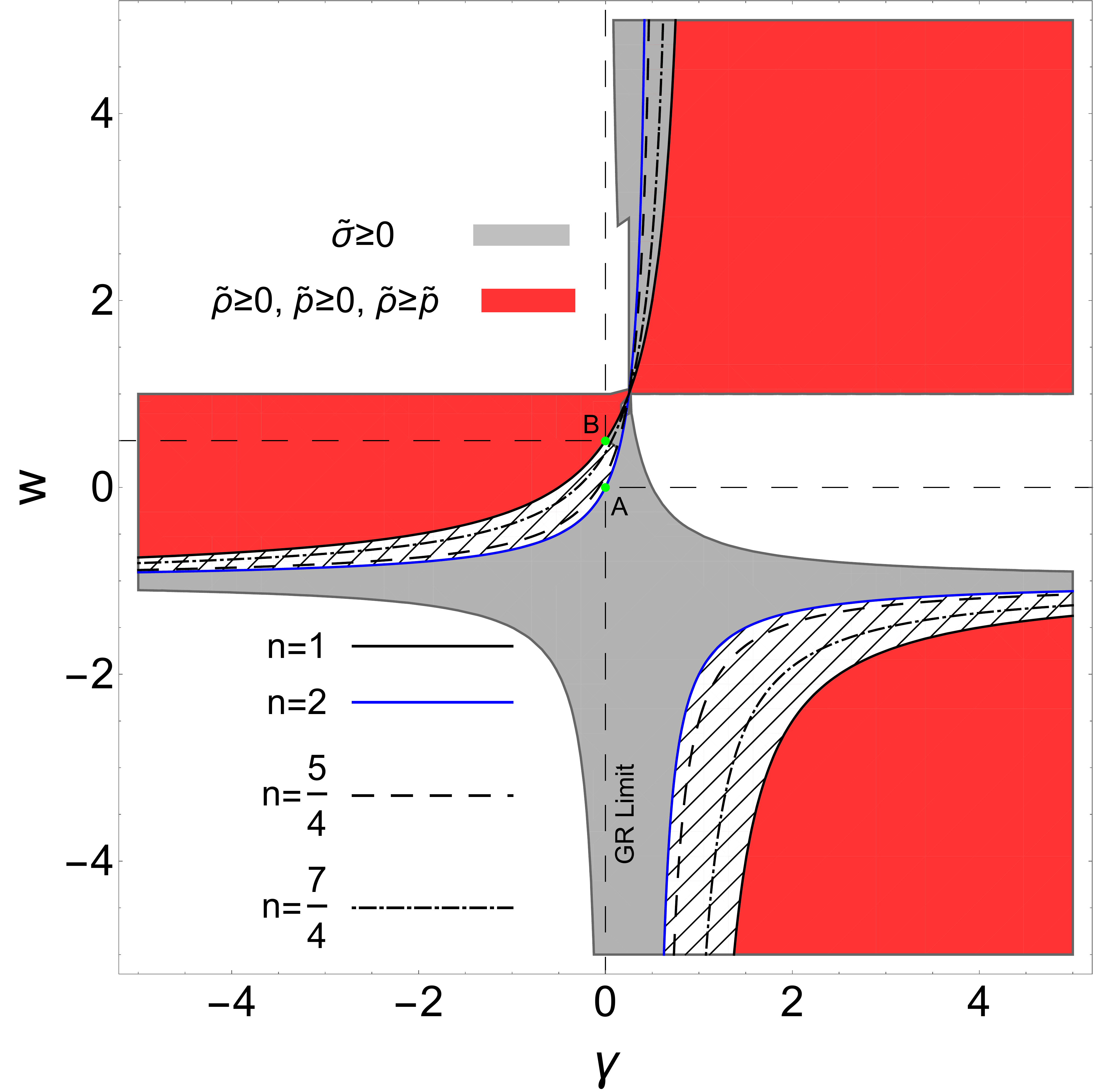}
		\caption{The allowed values of the Rastall parameter $\gamma$ and the EoS parameter $w$ that respect the ECs. The stripped region represents the intersection of gray and red regions for which all the ECs are fulfilled. The vertical dashed line represents the GR limit of the model, and the horizontal dashed lines display specific values of EoS parameter so that for the line segment $\overline{\sf AB}$ we have $0\leq w<\frac{1}{2}$.}\label{fig1}
	\end{center}
\end{figure}

For the existence of the outgoing radial null geodesics reaching the faraway observers, Eq.~(\ref{rooteq}) should admit real positive roots depending on the physically reasonable values of parameters $\{\gamma,w,\xi,\lambda\}$, generating a 4-dimensional space whose allowed regions are subject to fulfillment of the conditions \ref{I}-\ref{III}.
In particular, to satisfy condition \ref{I}, we demand that the pair $(\gamma,w)$ should fulfill the bounds given in Eq.~(\ref{ex1012-1}). We therefore have a two dimensional slice, as shown in Fig.~\ref{fig1}, which represents the domain of validity of energy conditions. For the red region, the second and third inequalities in Eq.~(\ref{wec}) and the second one in Eq.~(\ref{dec}) are satisfied while the gray one stands for validity of the first inequalities of both conditions. Therefore, the intersection of these two regions, as shown by the striped region, encompasses the condition  Eq.~(\ref{ex1012-1}).

In order to determine the degree $n(\gamma, w)$ (which is associated to $\beta(\gamma, w)$ through Eq.~(\ref{n-beta})) of the algebraic equation (\ref{rooteq}), we need to find suitable ranges of values for  $\{\gamma,w\}$ following the numerical analysis presented in Fig.~\ref{fig1}. The simplest choices for the exponent $n(\gamma, w)$ are $n=1$ and $n=2$ for which the corresponding  EoS parameters read, respectively
\begin{eqnarray}
w &=& \frac{2 \gamma }{1-2 \gamma}\, ,\quad  \,\, \,  (n=1),\\
w &=& \frac{1+2 \gamma}{2 (1-\gamma)}\, , \quad (n=2).
\label{EoSn12}
\end{eqnarray}
The values of parameters $\{\gamma, w\}$ satisfying the  equations above constitute two curves in Fig.~\ref{fig1} (the black solid curve for $n=1$, and the blue curve for $n=2$) that lay on the border of the striped region. 
Then,  to fulfill the condition \ref{I}, the allowed values of $n(\gamma, w)$ are all those curves that lie between the blue and the black solid curves. 
We note that for $n=2$ (or identically $\beta=0$), the parameter $\alpha(\gamma, w)$ (cf. Eq.~(\ref{beta})) is not defined, so we neglect this case\footnote{In fact, this case should be studied separately within the second solution of Eq.~(\ref{solmassf}) for the mass function. However, by using the chosen functions (\ref{F1v}) and (\ref{F2v}) this solution leads to an undefined value for the $\dot{m}$ in the limit where the singularity is approached. This implies that, in order to study the case $\beta=0$, we need to consider the free functions $F_1(v)$ and $F_2(v)$ different from those we have considered herein this work.}. We therefore require that $n$ holds the range $1\leq n(\gamma,w)<2$.

The case $n=1$ (or identically $\beta=1$), corresponds to a vanishing effective  pressure $\tilde{p}$ for the fluid (cf. Eq.~(\ref{effprofs})).
In this case, Eq.~(\ref{rooteq}) reduces to a quadratic equation:
\begin{eqnarray}\label{rooteqn1}
2\lambda Z_0^2+(2\xi-1)Z_0+2=0,
\end{eqnarray}
whose  solutions are
\begin{eqnarray}
Z_0^\pm=\f{1-2\xi\pm\left[1-16\lambda-4\xi+4\xi^2\right]^{\f{1}{2}}}{4\lambda}\, .
\label{solreqn1}
\end{eqnarray}
By substituting the  solutions above into the Eq. (\ref{slopeah}) we find the slope of the apparent horizon  as
\begin{eqnarray}
X_0 =\frac{1-2 \xi }{2 \lambda }.
\label{solreqn1aph}
\end{eqnarray}
Then, the  conditions \ref{II} and \ref{III} demand the following restrictions on the pair $\{\lambda,\xi\}$:
\begin{eqnarray}
0<\xi <\frac{1}{2}\, , \quad {\rm and} \quad  0<\lambda \leq \frac{1}{16} \left(4 \xi ^2-4 \xi +1\right). \quad 
\label{II-III}
\end{eqnarray}
The solution (\ref{solreqn1}) looks similar  to the solution presented in Ref.~\cite{Mkenyeleye:2014dwa} for a general relativistic model (being identical to our model with $\gamma=0$). However, unlike their fluid model with a zero pressure,  in our  model (i.e., for $\gamma\neq0$) only the effective pressure  $\tilde{p}$ vanishes while the fluid itself can contain non-vanishing pressures. This is a consequence of the mutual interaction between spacetime and matter which shows itself as the non-minimally coupled term within the field equations with Rastall parameter $\gamma$ being the strength of such coupling.

\par
\begin{figure*}
	\begin{center}
		\includegraphics[scale=0.3]{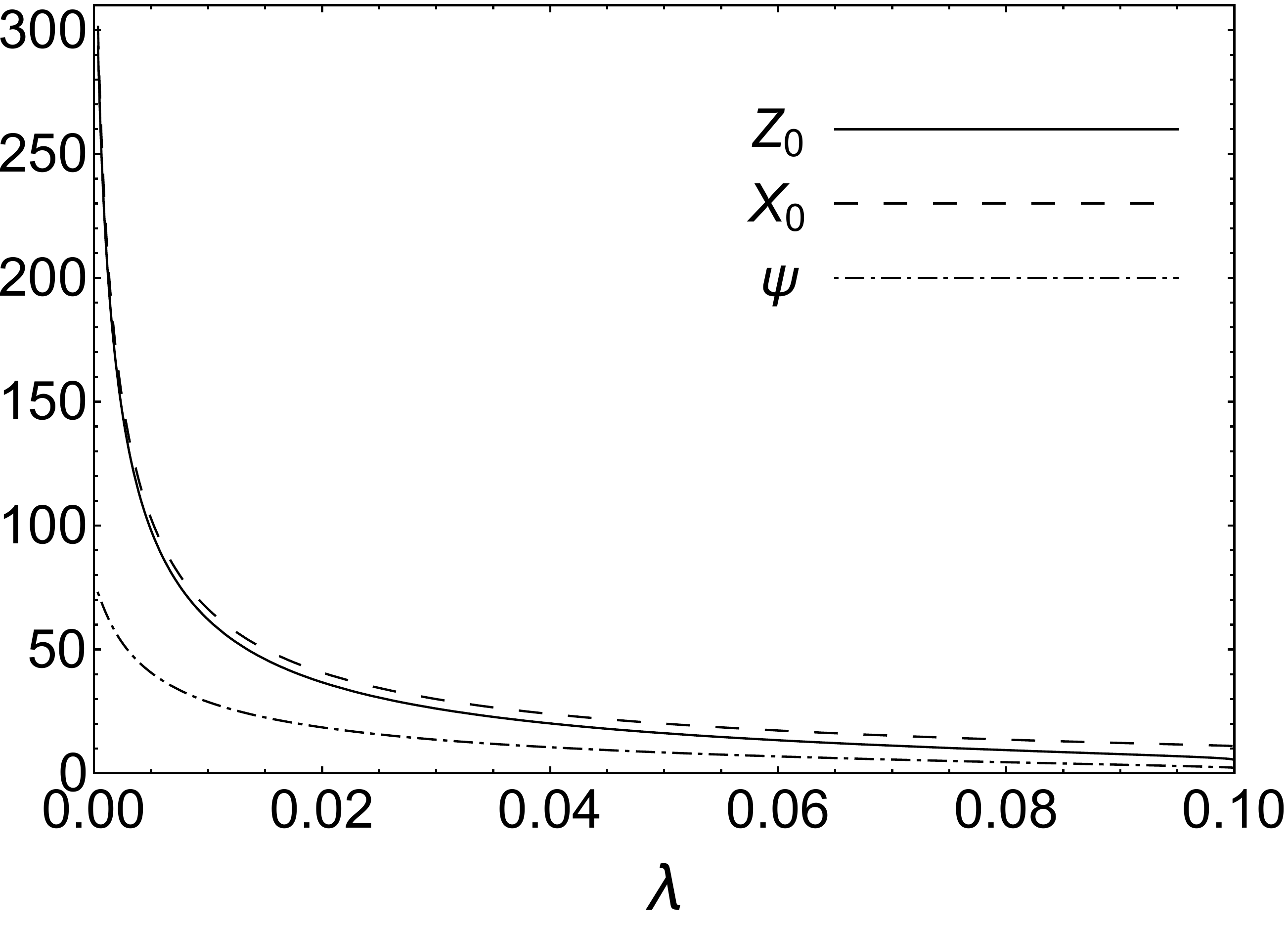} \quad \quad  
		\includegraphics[scale=0.26]{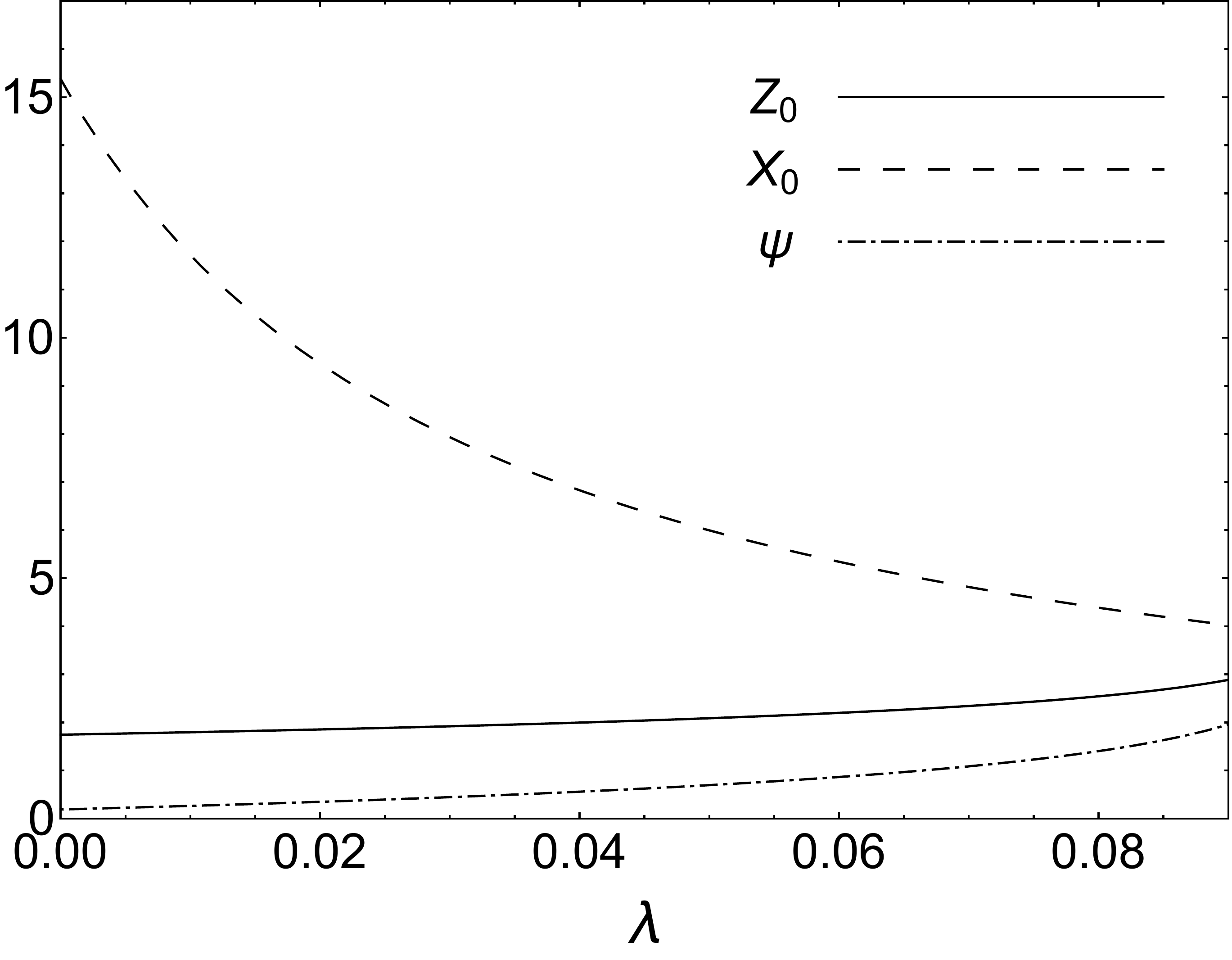}
		\caption{Behaviors of the tangents to the radial null geodesics (i.e. $Z_0$) and the apparent horizons (i.e. $X_0$), and the parameter $\psi$, in terms of the parameter $\lambda$. Left panel:  for the choice of values $\xi=0.099$ and $n=5/4$. Right panel:  for the choice of values $\xi=0.012$ and $n=7/4$. In both plots, we have $Z_0<X_0$ and $\psi>0$.}\label{fig2}
	\end{center}
\end{figure*}

Let us now study other possible solutions to the equation (\ref{rooteq}) for the cases in which $1<n(\gamma, w)<2$. As an example, let us consider the case $n=5/4$. The EoS then reads
\begin{eqnarray}
w=-\frac{1+14 \gamma}{2 (7 \gamma -4)}\, .
\label{n54}
\end{eqnarray}
In Fig.~\ref{fig1} we have plotted the  EoS parameter as a function of Rastall parameter (see the black dashed curve in the striped region). We observe that some values of $\{\gamma,w\}$  lie within the striped region for which the condition \ref{I} is satisfied. Now Eq.~(\ref{rooteq}) can be re-expressed as
\begin{eqnarray}\label{root54}
8 \xi{Z_0}^{5/4}+6 \lambda{Z_0}^2-3 {Z_0}+6=0.
\end{eqnarray}
The  equation above cannot be solved analytically; thus, we proceed with finding the roots numerically. By doing so, we arrive at a three-dimensional parameter space constructed by $\left(Z_0(\lambda,\xi),\lambda,\xi\right)$. The left panel in Fig.~\ref{fig2} represents the numerical solutions for Eq.~(\ref{root54})  (black solid curve) along with the expression (\ref{slopeah}) (dashed curve) in terms of the parameter $\gamma$. We observe that $0<Z_0<X_0$ so the conditions \ref{II} and \ref{III} are satisfied which implies that the central singularity will be naked as the collapse final state.

As another example, let us consider the case where $n=7/4$. The EoS for this case reads
\begin{eqnarray}\label{n74}
w=-\frac{3+10 \gamma}{2 (5 \gamma -4)},
\end{eqnarray}
which corresponds to the black dot-dashed curve in Fig.~\ref{fig1}. The  Eq.~(\ref{rooteq}) for $n=7/4$  can be rewritten then:
\begin{eqnarray}\label{root74}
8 \xi{Z_0}^{7/4}+2\lambda{Z_0}^2-{Z_0}+2=0.
\end{eqnarray}
For a fixed value of $\xi>0$, numerical solution to the  equation above reveals that the tangent to the radial null geodesics (black curve) is positive for the ranges of parameter $\lambda>0$ (cf.  the right panel in Fig.~\ref{fig2}). Moreover, this tangent is smaller than the slope of apparent horizon (dashed curve) in the limit of singular node, hence, the conditions \ref{II} and \ref{III} are fulfilled.  We further note that for all the  cases above we have  $\lambda>0$ and $\xi>0$ for $F_1(v)$ and $M(v)$ to be positive, increasing functions of the advanced time. Therefore, in order that $\psi>0$ in Eq. (\ref{Psi}), the condition $0<\beta(\gamma, w)\leq1$ must be fulfilled. Clearly this happens for the range of the exponent $1\leq n(\gamma, w)<2$ for which  the condition \ref{IV} (cf. Eq.~(\ref{curvstr})) is satisfied (cf. the dot-dashed curve within the both panels of Fig.~\ref{fig2}). Thus the central singularity is gravitationally strong, and hence we are led to the physically reasonable solutions for the naked singularity formation as the collapse end state in our herein model.
\par

Another interesting feature of the present  model  that begs more consideration is its general relativistic limit (i.e. when $\gamma=0$). For this case we get the components of Eq.~(\ref{EMTnm}) in the limit $\gamma\rightarrow 0$ as
\begin{eqnarray}\label{GRL}
\sigma\big|_{\gamma\to0}=\sigma_0=\frac{2\dot{M}}{r^2}+\frac{2\dot{F_1}}{1-2w}r^{-(1+2w)},~~~~~\rho\big|_{\gamma\to0}=\rho_0=2F_1r^{-2(1+w)},~~~~~p\big|_{\gamma\to0}=p_0=2wF_1r^{-2(1+w)}.
\end{eqnarray}
By imposing the ECs (\ref{wec}) and (\ref{dec}) into the second and third relations of the  equation above we get a constraint on $w$ as $0\leq w\leq1$. In addition, by considering the free functions $F_1(v)$ and $M(v)$ as defined in Eqs.~(\ref{F1v}) and (\ref{F2v}), we get the density of the pure radiation field as
\begin{equation}
\sigma_0=\frac{2\lambda}{r^2}+\frac{2\xi w}{r^2(1-2w)}\left(\frac{r}{v}\right)^{1-2w}.
\end{equation}
Since $\lambda>0$ and $\xi>0$, the condition $\sigma_0\geq0$ requires that $w<1/2$. Thus, all the ECs are satisfied for $0\leq w<1/2$. This allowed range of the EoS parameter $w$ is presented by the line $\overline{\sf AB}$ connecting the green points in Fig.~\ref{fig1}. We therefore observe that in Rastall theory, there exists a wider range for  $w$  with respect to GR (i.e. $-1<w<1/2$ for $\gamma<0$ and $w<-1$ for $\gamma>0$)  due to which the gravitational collapse of a physically reasonable null fluid can end up with formation of a naked singularity. Since the Rastall parameter can be interpreted as a measure of non-conservation of the EMT, or in view of Eq.~(\ref{T-eff}) as the strength of the mutual interaction between the geometry and matter, one can intuitively imagine that such a non-conservation  goes in favor of a naked singularity formation  rather than a black hole formation.

\begin{figure*}
	\begin{center}
		\includegraphics[scale=0.23]{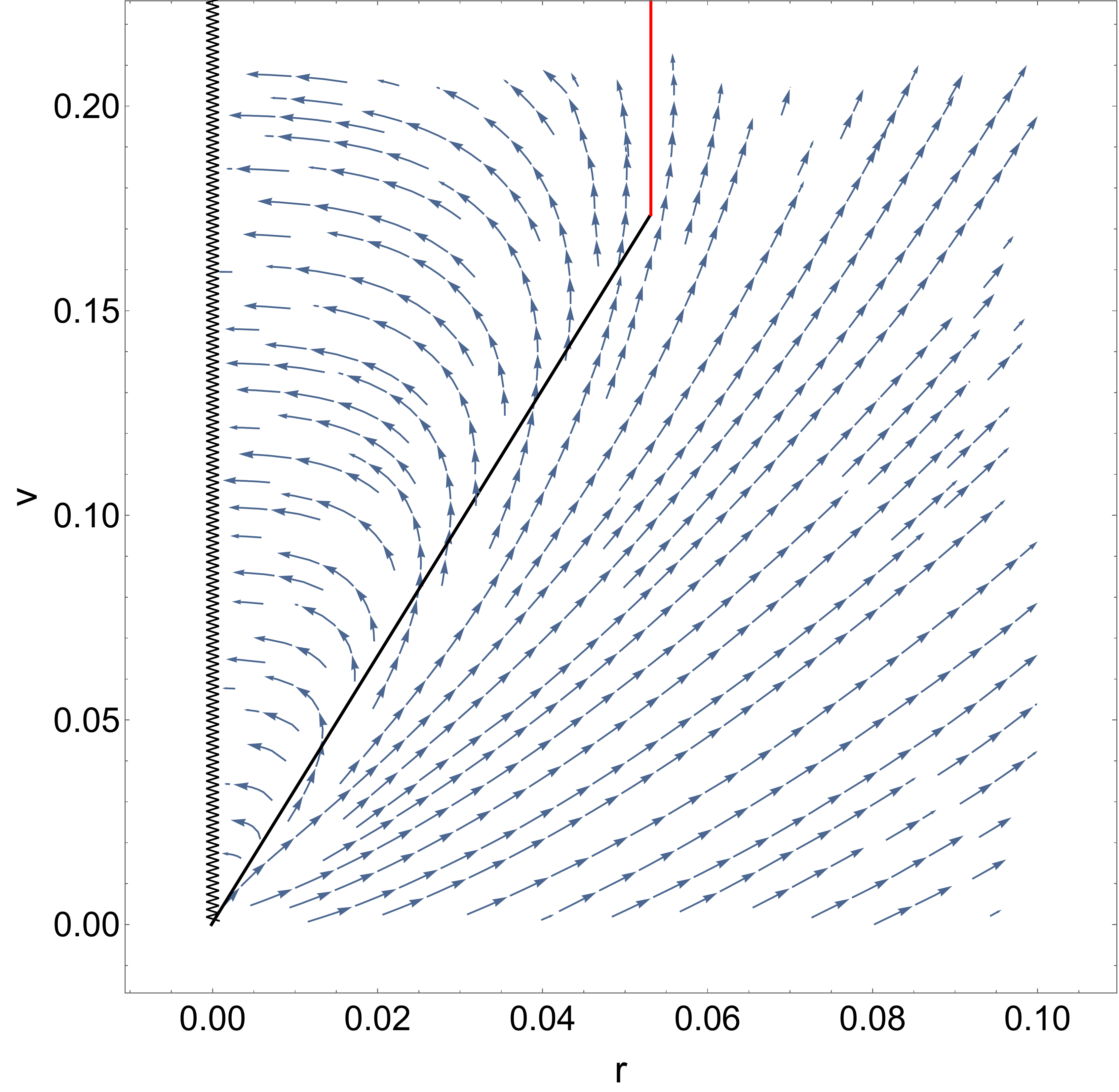} \quad \quad  
		\caption{Behavior of radial null geodesics in the vicinity of the singular region (jagged line) for $\xi=0.099$ and $\lambda=0.091$ and $C_0=10000$. The black solid line represents the apparent horizon curve the vertical red line indicates the event horizon. We have chosen the values of the pair $(\gamma,w)$ from the dashed curve of Fig. (\ref{fig2}).}\label{fig3}
	\end{center}
\end{figure*}

\section{Concluding Remarks}
\label{Conclusion}

In the present work we studied gravitational collapse of a type II fluid in the framework of Rastall gravity. By considering a particular choice for the Vaidya mass $m(r,v)$, we observed that depending on the model parameters $\{\gamma, w, \xi, \lambda\}$, {\em strong} curvature {\em naked singularities} could emerge as the collapse final states. The nakedness of these singularities were examined by pursuing the radial null geodesics terminating in the past at the central singularities with positive tangents to the geodesic curves, $(dv/dr)_{v,r=0}$ (cf. Eqs.~(\ref{Z0solve}) and (\ref{rooteq})). However, the existence of such geodesic curves could not necessarily imply that the singularities are naked, as these curves might be turned back to the singularity from their starting points due to formation of apparent horizons. To deal with this issue, we computed the tangent to the apparent horizon, $(dv/dr)_{\rm AH}$ (cf. Eq.~(\ref{slopeah})), in the limit where the singularity is reached. 
It was observed that for the certain values of the model parameters, the tangents to the apparent horizons are larger than those of the geodesics, hence, the radial null geodesics could emerge from the singularities standing outside the trapped regions so that they could arrive in the exterior universe  by remaining untrapped.
 
By fixing the  conditions above, the solutions generating naked singularity  were classified through the parameter $n(\gamma, w)$ whose values were subject to the fulfillment of the ECs. A particular  solution (i.e. the case $n=1$) was found for which the collapse scenario ends in a naked singularity 
whose null fluid matter field displayed an effective dust-like behaviour. Despite the vanishing effective pressure profile (induced  by Rastall gravity modifications), the fluid itself had nonzero pressure. 
This  is a consequence of the non-minimal coupling between  geometry and matter through the Rastall  parameter $\gamma\neq0$.

As discussed in Sec.~\ref{localnaked}, the Vaidya mass we considered in this work was an increment mass function of the coordinates $(v, r)$ initiated from an empty Minkowski spacetime  (see  Eq.~(\ref{solmassf-2}). Its arbitrary functions were chosen as $F_1(v)=\xi v^{1-\beta}$ and $M(v)=\lambda v$ (cf. \ref{F1v}) and (\ref{F2v}). However, if we chose instead the two different free functions as $F_1(v)=a_0/2$ and $M(v)=0$ with $a_0$ being a constant, then the mass function would become $m(r)=(a_0/2)r$, with the effective density profiles:
\begin{eqnarray}
\tilde{\rho} = \f{a_0}{\kappa r^2},  \quad  \tilde{p}=\sigma =0, \quad
p=\f{2\gamma}{1-2\gamma}\rho=\f{2a_0\gamma}{\kappa r^2}. \quad\quad 
\end{eqnarray}
This solution represents the gravitational field of a monopole~\cite{Barriola:1989hx}. The gravitational collapse of a monopole and the situation  where naked singularities can form have been studied in Refs.~\cite{Joshi:1987wg,Joshi:2008zz}. It is also remarkable that the static limit of the solution (\ref{solmassf}) is obtained by setting the free functions to be constants as $F_1(v)=F_{0}$ and $M(v)=M_{0}$. Then, by using  the coordinate transformation $dt=dv-dr/(1-2m(r)/r)$ one can recast the metric (\ref{vaidyametr}) into the Schwarzschild coordinates as 
\begin{align}
ds^2=-f(r)dt^2+\f{dr^2}{f(r)}+r^2d\Omega^2,~~~~~~~~~~f(r)=1-\f{2M_0}{r}-\f{2\alpha F_0}{r^{\beta-1}}.
\label{Kiselev-like}
\end{align}
By comparing  the  solution (\ref{Kiselev-like}) with the results of Ref.~\cite{Heydarzade_2017}  we observe that for the choice of parameter $w=(1+3\omega_s)/2$, as a relation between the {\rm EoS} of the two-fluids system and the barotropic {\rm EoS} for the surrounding field, within the herein Rastall gravity inspired collapse, the uncharged Kiselev-like black hole solutions surrounded by a perfect fluid  are recovered.

\par
Finally, we examine the local versus   global nakedness of the spacetime singularity. By considering the solution (\ref{solmassf-2}), we find that the spacetime is {\em self-similar}  admitting a homothetic Killing vector field \cite{Joshi:1987wg}:
\begin{equation}\label{HomKill}
{\sf K}^\mu=r\frac{\partial}{\partial r}+v\frac{\partial}{\partial v},
\end{equation}
which satisfies the condition $\pounds_{\sf K} g_{\mu\nu}=2g_{\mu\nu}$. 
We can therefore define a conserved quantity along the radial
null geodesics as
\begin{align}\label{HomKill1}
r\zeta_r+v\zeta_v = {\rm const}.
\end{align}
Thereby, we get the components of $\zeta^\mu$ as
\begin{align}\label{zetacomps}
\zeta^v=\frac{2C_0r}{2r^2-{v}\left(r-{2m}\right)},~~~~~~~~~~~\zeta^r=\frac{C_0(r-2m)}{r(2r-v)+2mv},
\end{align}
where the constant of motion is denoted by $C_0$. Fig.~\ref{fig3} represents the numerical plot of  behavior of the vector fields $(\zeta^v,\zeta^r)$ within the $v-r$ plane where the black solid line displays the location of the points  satisfying the apparent horizon equation (\ref{AHEQ}). The slope of this line is given by Eq.~(\ref{diffapphorcurve}). The apparent horizon develops to meet the red line at which $dv/dr\rightarrow\infty$. 
This line through which the outgoing null geodesics can  never escape, displays the location of the event horizon that forms for $v>v_0$ and $r=2m(r,v)=2M_0={\rm constant}$. It should be noted that for $v>v_0$ all the points to the left of the red line are out of the  causal contact with the future null infinity ($\scri^+$) while the points to the right are. We can find a congruence of outgoing light rays emanating from the singular point at $(r,v)=(0,0)$ with an initial slope being smaller than that of the apparent horizon. However, these rays will be captured later on by the pull of gravity and will fall into the curvature singularity, hence, the singularity in this case would be visible only to its neighboring observes. In other words, the singularity in this case is locally naked and the strong version of CCC is violated. We further observe that there exist families of radial null geodesics which emerge from the singularity with $Z_0<X_0$ and will escape to $\scri^+$. In this case, the spacetime possesses a globally naked singularity, i.e., an ultra-dense region of the extreme gravity can be detected by faraway observers and consequently the weak version of CCC is violated. We therefore conclude that Rastall theory of gravity can also provide a framework which leads to the remarkable solutions for occurrence of strong curvature naked singularities being counterexamples to the {\rm CCC}.

As final remarks, we would  mention that the collapse of a homogeneous perfect fluid in Rastall gravity has been studied earlier in Ref.~\cite{Ziaie_2019} where it was shown that both naked singularities and black holes can form as the collapse outcome but for  different values of the model parameters from those of GR. However, our attempt in the present work was to explore other features of the collapse process in Rastall gravity by considering a different type of matter field from those assumed  in Ref.~\cite{Ziaie_2019}. We obtained a class of spacetimes  with a generalized Vaidya metric  describing a spherically symmetric  collapse of a combination of Type I and Type II matter fields in an initially empty region of Minkowski spacetime.
We further found that both locally and globally naked singularities can be born as the collapse outcome for a wider range of model parameters in comparison to those of GR. These solutions can be generalized further to the cases of NSQF~\cite{Harko:2000ni,Ghosh_2003} as well as the higher dimensional Vaidya spacetimes~\cite{Ghosh:2008zza}. In the former, one may consider the EoS of strange quark matter, i.e. $p=w\rho+b$, where $b$ is a  parameter related to the bag constant \cite{Witten:1984rs,weinberg_1996}.  In the GR framework, the bag constant contributes to the generalized Vaidya mass similar to a cosmological constant \cite{Harko:2000ni,Ghosh_2003}. Hence, exploring these two cases can be of interest in view of possible impact of mutual matter-geometry interaction on the final fate of gravitational collapse of NSQF and also the relation between such an interaction with higher-dimensions. The results of future investigations on these issues will be reported as an independent work.

\acknowledgments{The authors would like to sincerely appreciate the anonymous reviewers for constructive comments and suggestions that helped us to improve the original version of the manuscript. This article is based upon work from European Cooperation in Science and Technology (COST) action CA18108--Quantum gravity phenomenology in the multimessenger approach--supported by COST. }

\bibliography{Bibliography}

\begin{thebibliography}{123}
\expandafter\ifx\csname natexlab\endcsname\relax\def\natexlab#1{#1}\fi
\expandafter\ifx\csname bibnamefont\endcsname\relax
  \def\bibnamefont#1{#1}\fi
\expandafter\ifx\csname bibfnamefont\endcsname\relax
  \def\bibfnamefont#1{#1}\fi
\expandafter\ifx\csname citenamefont\endcsname\relax
  \def\citenamefont#1{#1}\fi
\expandafter\ifx\csname url\endcsname\relax
  \def\url#1{\texttt{#1}}\fi
\expandafter\ifx\csname urlprefix\endcsname\relax\def\urlprefix{URL }\fi
\providecommand{\bibinfo}[2]{#2}
\providecommand{\eprint}[2][]{\url{#2}}

\bibitem[{\citenamefont{Hawking and Penrose}(1996)}]{Hawking:1996jh}
\bibinfo{author}{\bibfnamefont{S.}~\bibnamefont{Hawking}} \bibnamefont{and}
  \bibinfo{author}{\bibfnamefont{R.}~\bibnamefont{Penrose}},
  \emph{\bibinfo{title}{{The Nature of space and time}}}
  (\bibinfo{publisher}{Princeton, USA: Univ. Pr.}, \bibinfo{year}{1996}).

\bibitem[{\citenamefont{Hawking and Ellis}(2011)}]{Hawking:1973uf}
\bibinfo{author}{\bibfnamefont{S.~W.} \bibnamefont{Hawking}} \bibnamefont{and}
  \bibinfo{author}{\bibfnamefont{G.~F.~R.} \bibnamefont{Ellis}},
  \emph{\bibinfo{title}{{The Large Scale Structure of Space-Time}}}
  (\bibinfo{publisher}{Cambridge University Press}, \bibinfo{year}{2011}).

\bibitem[{\citenamefont{Penrose}(1969)}]{Penrose:1969pc}
\bibinfo{author}{\bibfnamefont{R.}~\bibnamefont{Penrose}},
  \bibinfo{journal}{Riv. Nuovo Cim.} \textbf{\bibinfo{volume}{1}},
  \bibinfo{pages}{252} (\bibinfo{year}{1969}), \bibinfo{note}{[Gen. Rel.
  Grav.34,1141(2002)]}.

\bibitem[{\citenamefont{Wald}(1998)}]{wald1998black}
\bibinfo{author}{\bibfnamefont{R.}~\bibnamefont{Wald}},
  \emph{\bibinfo{title}{Black Holes and Relativistic Stars}}
  (\bibinfo{publisher}{University of Chicago Press}, \bibinfo{year}{1998}),
  ISBN \bibinfo{isbn}{9780226870359}.

\bibitem[{\citenamefont{Wald}(1984)}]{wald1984general}
\bibinfo{author}{\bibfnamefont{R.}~\bibnamefont{Wald}},
  \emph{\bibinfo{title}{General Relativity}} (\bibinfo{publisher}{University of
  Chicago Press}, \bibinfo{year}{1984}), ISBN \bibinfo{isbn}{9780226870328}.

\bibitem[{\citenamefont{Clarke}(1994)}]{Clarke:1994}
\bibinfo{author}{\bibfnamefont{C.~J.~S.} \bibnamefont{Clarke}},
  \bibinfo{journal}{Class.Quant.Grav.} \textbf{\bibinfo{volume}{11}},
  \bibinfo{pages}{1375} (\bibinfo{year}{1994}).

\bibitem[{\citenamefont{Wald}(1997)}]{Wald:1997wa}
\bibinfo{author}{\bibfnamefont{R.~M.} \bibnamefont{Wald}}, in
  \emph{\bibinfo{booktitle}{{Black Holes, Gravitational Radiation and the
  Universe: Essays in Honor of C.V. Vishveshwara}}} (\bibinfo{year}{1997}), pp.
  \bibinfo{pages}{69--85}, \eprint{gr-qc/9710068}.

\bibitem[{\citenamefont{Earman}(1995)}]{earman1995bangs}
\bibinfo{author}{\bibfnamefont{J.}~\bibnamefont{Earman}},
  \emph{\bibinfo{title}{Bangs, Crunches, Whimpers, and Shrieks: Singularities
  and Acausalities in Relativistic Spacetimes}} (\bibinfo{publisher}{Oxford
  University Press}, \bibinfo{year}{1995}), ISBN \bibinfo{isbn}{9780195344646}.

\bibitem[{\citenamefont{Joshi}(2014)}]{Joshi:2013xoa}
\bibinfo{author}{\bibfnamefont{P.~S.} \bibnamefont{Joshi}},
  \emph{\bibinfo{title}{{Spacetime Singularities}}} (\bibinfo{year}{2014}), pp.
  \bibinfo{pages}{409--436}, \eprint{1311.0449}.

\bibitem[{\citenamefont{Joshi}(1987)}]{Joshi:1987wg}
\bibinfo{author}{\bibfnamefont{P.~S.} \bibnamefont{Joshi}},
  \emph{\bibinfo{title}{{Global aspects in gravitation and cosmology}}}
  (\bibinfo{publisher}{Clarendon Press}, \bibinfo{year}{1987}).

\bibitem[{\citenamefont{Joshi}(2012)}]{Joshi:2008zz}
\bibinfo{editor}{\bibfnamefont{P.~S.} \bibnamefont{Joshi}}, ed.,
  \emph{\bibinfo{title}{{Gravitational Collapse and Spacetime Singularities}}},
  Cambridge Monographs on Mathematical Physics (\bibinfo{publisher}{Cambridge
  University Press}, \bibinfo{year}{2012}), ISBN \bibinfo{isbn}{9781107405363,
  9780521871044, 9780511372834}.

\bibitem[{\citenamefont{Joshi and Dwivedi}(1993)}]{Joshi:1993zg}
\bibinfo{author}{\bibfnamefont{P.~S.} \bibnamefont{Joshi}} \bibnamefont{and}
  \bibinfo{author}{\bibfnamefont{I.~H.} \bibnamefont{Dwivedi}},
  \bibinfo{journal}{Phys. Rev.} \textbf{\bibinfo{volume}{D47}},
  \bibinfo{pages}{5357} (\bibinfo{year}{1993}), \eprint{gr-qc/9303037}.

\bibitem[{\citenamefont{Maeda et~al.}(1998)\citenamefont{Maeda, Torii, and
  Narita}}]{Maeda:1998hs}
\bibinfo{author}{\bibfnamefont{K.}~\bibnamefont{Maeda}},
  \bibinfo{author}{\bibfnamefont{T.}~\bibnamefont{Torii}}, \bibnamefont{and}
  \bibinfo{author}{\bibfnamefont{M.}~\bibnamefont{Narita}},
  \bibinfo{journal}{Phys. Rev. Lett.} \textbf{\bibinfo{volume}{81}},
  \bibinfo{pages}{5270} (\bibinfo{year}{1998}), \eprint{gr-qc/9810081}.

\bibitem[{\citenamefont{Giambo}(2005)}]{Giambo:2005se}
\bibinfo{author}{\bibfnamefont{R.}~\bibnamefont{Giambo}},
  \bibinfo{journal}{Class. Quant. Grav.} \textbf{\bibinfo{volume}{22}},
  \bibinfo{pages}{2295} (\bibinfo{year}{2005}), \eprint{gr-qc/0501013}.

\bibitem[{\citenamefont{Bhattacharya et~al.}(2011)\citenamefont{Bhattacharya,
  Goswami, and Joshi}}]{Bhattacharya:2010tc}
\bibinfo{author}{\bibfnamefont{S.}~\bibnamefont{Bhattacharya}},
  \bibinfo{author}{\bibfnamefont{R.}~\bibnamefont{Goswami}}, \bibnamefont{and}
  \bibinfo{author}{\bibfnamefont{P.~S.} \bibnamefont{Joshi}},
  \bibinfo{journal}{Int. J. Mod. Phys.} \textbf{\bibinfo{volume}{D20}},
  \bibinfo{pages}{1123} (\bibinfo{year}{2011}), \eprint{1010.1757}.

\bibitem[{\citenamefont{Tavakoli
  et~al.}(2013{\natexlab{a}})\citenamefont{Tavakoli, Marto, Ziaie, and
  Vargas~Moniz}}]{Tavakoli:2011iq}
\bibinfo{author}{\bibfnamefont{Y.}~\bibnamefont{Tavakoli}},
  \bibinfo{author}{\bibfnamefont{J.}~\bibnamefont{Marto}},
  \bibinfo{author}{\bibfnamefont{A.~H.} \bibnamefont{Ziaie}}, \bibnamefont{and}
  \bibinfo{author}{\bibfnamefont{P.}~\bibnamefont{Vargas~Moniz}},
  \bibinfo{journal}{Gen. Rel. Grav.} \textbf{\bibinfo{volume}{45}},
  \bibinfo{pages}{819} (\bibinfo{year}{2013}{\natexlab{a}}),
  \eprint{1105.0445}.

\bibitem[{\citenamefont{Nakonieczna et~al.}(2015)\citenamefont{Nakonieczna,
  Rogatko, and Nakonieczny}}]{Nakonieczna_2015}
\bibinfo{author}{\bibfnamefont{A.}~\bibnamefont{Nakonieczna}},
  \bibinfo{author}{\bibfnamefont{M.}~\bibnamefont{Rogatko}}, \bibnamefont{and}
  \bibinfo{author}{\bibfnamefont{u.}~\bibnamefont{Nakonieczny}},
  \bibinfo{journal}{Journal of High Energy Physics}
  \textbf{\bibinfo{volume}{2015}} (\bibinfo{year}{2015}).

\bibitem[{\citenamefont{Giambò}(2009)}]{Giamb__2009}
\bibinfo{author}{\bibfnamefont{R.}~\bibnamefont{Giambò}},
  \bibinfo{journal}{Journal of Mathematical Physics}
  \textbf{\bibinfo{volume}{50}}, \bibinfo{pages}{012501}
  (\bibinfo{year}{2009}).

\bibitem[{\citenamefont{Giambo et~al.}(2008)\citenamefont{Giambo, Giannoni, and
  Magli}}]{Giambo:2008ya}
\bibinfo{author}{\bibfnamefont{R.}~\bibnamefont{Giambo}},
  \bibinfo{author}{\bibfnamefont{F.}~\bibnamefont{Giannoni}}, \bibnamefont{and}
  \bibinfo{author}{\bibfnamefont{G.}~\bibnamefont{Magli}}, \bibinfo{journal}{J.
  Math. Phys.} \textbf{\bibinfo{volume}{49}}, \bibinfo{pages}{042504}
  (\bibinfo{year}{2008}), \eprint{0802.0992}.

\bibitem[{\citenamefont{Ghosh and Dadhich}(2002)}]{Ghosh:2002mf}
\bibinfo{author}{\bibfnamefont{S.~G.} \bibnamefont{Ghosh}} \bibnamefont{and}
  \bibinfo{author}{\bibfnamefont{N.}~\bibnamefont{Dadhich}},
  \bibinfo{journal}{Phys. Rev.} \textbf{\bibinfo{volume}{D65}},
  \bibinfo{pages}{127502} (\bibinfo{year}{2002}), \eprint{gr-qc/0204091}.

\bibitem[{\citenamefont{Goswami and Joshi}(2007)}]{Goswami_2007}
\bibinfo{author}{\bibfnamefont{R.}~\bibnamefont{Goswami}} \bibnamefont{and}
  \bibinfo{author}{\bibfnamefont{P.~S.} \bibnamefont{Joshi}},
  \bibinfo{journal}{Physical Review D} \textbf{\bibinfo{volume}{76}}
  (\bibinfo{year}{2007}).

\bibitem[{\citenamefont{Giambò and Quintavalle}(2008)}]{Giamb__2008}
\bibinfo{author}{\bibfnamefont{R.}~\bibnamefont{Giambò}} \bibnamefont{and}
  \bibinfo{author}{\bibfnamefont{S.}~\bibnamefont{Quintavalle}},
  \bibinfo{journal}{Classical and Quantum Gravity}
  \textbf{\bibinfo{volume}{25}}, \bibinfo{pages}{145003}
  (\bibinfo{year}{2008}).

\bibitem[{\citenamefont{Yamada and Shinkai}(2011)}]{Yamada_2011}
\bibinfo{author}{\bibfnamefont{Y.}~\bibnamefont{Yamada}} \bibnamefont{and}
  \bibinfo{author}{\bibfnamefont{H.-a.} \bibnamefont{Shinkai}},
  \bibinfo{journal}{Physical Review D} \textbf{\bibinfo{volume}{83}}
  (\bibinfo{year}{2011}), ISSN \bibinfo{issn}{1550-2368}.

\bibitem[{\citenamefont{Dadhich
  et~al.}(2013{\natexlab{a}})\citenamefont{Dadhich, Ghosh, and
  Jhingan}}]{Dadhich_2013}
\bibinfo{author}{\bibfnamefont{N.}~\bibnamefont{Dadhich}},
  \bibinfo{author}{\bibfnamefont{S.~G.} \bibnamefont{Ghosh}}, \bibnamefont{and}
  \bibinfo{author}{\bibfnamefont{S.}~\bibnamefont{Jhingan}},
  \bibinfo{journal}{Physical Review D} \textbf{\bibinfo{volume}{88}}
  (\bibinfo{year}{2013}{\natexlab{a}}).

\bibitem[{\citenamefont{Shimano and Miyamoto}(2014)}]{Shimano_2014}
\bibinfo{author}{\bibfnamefont{M.}~\bibnamefont{Shimano}} \bibnamefont{and}
  \bibinfo{author}{\bibfnamefont{U.}~\bibnamefont{Miyamoto}},
  \bibinfo{journal}{Classical and Quantum Gravity}
  \textbf{\bibinfo{volume}{31}}, \bibinfo{pages}{045002}
  (\bibinfo{year}{2014}).

\bibitem[{\citenamefont{Rahman et~al.}(2019)\citenamefont{Rahman, Chakraborty,
  SenGupta, and Sen}}]{Rahman:2018oso}
\bibinfo{author}{\bibfnamefont{M.}~\bibnamefont{Rahman}},
  \bibinfo{author}{\bibfnamefont{S.}~\bibnamefont{Chakraborty}},
  \bibinfo{author}{\bibfnamefont{S.}~\bibnamefont{SenGupta}}, \bibnamefont{and}
  \bibinfo{author}{\bibfnamefont{A.~A.} \bibnamefont{Sen}},
  \bibinfo{journal}{JHEP} \textbf{\bibinfo{volume}{03}}, \bibinfo{pages}{178}
  (\bibinfo{year}{2019}), \eprint{1811.08538}.

\bibitem[{\citenamefont{Moore}(2006)}]{moore2006trends}
\bibinfo{author}{\bibfnamefont{D.}~\bibnamefont{Moore}},
  \emph{\bibinfo{title}{Trends in Quantum Gravity Research}}
  (\bibinfo{publisher}{Nova Science Publishers}, \bibinfo{year}{2006}), ISBN
  \bibinfo{isbn}{9781594546709}.

\bibitem[{\citenamefont{Bojowald}(2007)}]{Bojowald:2007ky}
\bibinfo{author}{\bibfnamefont{M.}~\bibnamefont{Bojowald}},
  \bibinfo{journal}{AIP Conf. Proc.} \textbf{\bibinfo{volume}{910}},
  \bibinfo{pages}{294} (\bibinfo{year}{2007}), \eprint{gr-qc/0702144}.

\bibitem[{\citenamefont{Tavakoli
  et~al.}(2013{\natexlab{b}})\citenamefont{Tavakoli, Marto, Ziaie, and
  Vargas~Moniz}}]{Tavakoli:2013tpa}
\bibinfo{author}{\bibfnamefont{Y.}~\bibnamefont{Tavakoli}},
  \bibinfo{author}{\bibfnamefont{J.}~\bibnamefont{Marto}},
  \bibinfo{author}{\bibfnamefont{A.~H.} \bibnamefont{Ziaie}}, \bibnamefont{and}
  \bibinfo{author}{\bibfnamefont{P.}~\bibnamefont{Vargas~Moniz}},
  \bibinfo{journal}{Phys. Rev. D} \textbf{\bibinfo{volume}{87}},
  \bibinfo{pages}{024042} (\bibinfo{year}{2013}{\natexlab{b}}).

\bibitem[{\citenamefont{Marto et~al.}(2015)\citenamefont{Marto, Tavakoli, and
  Vargas~Moniz}}]{Marto:2013soa}
\bibinfo{author}{\bibfnamefont{J.}~\bibnamefont{Marto}},
  \bibinfo{author}{\bibfnamefont{Y.}~\bibnamefont{Tavakoli}}, \bibnamefont{and}
  \bibinfo{author}{\bibfnamefont{P.}~\bibnamefont{Vargas~Moniz}},
  \bibinfo{journal}{Int. J. Mod. Phys. D} \textbf{\bibinfo{volume}{24}},
  \bibinfo{pages}{1550025} (\bibinfo{year}{2015}), \eprint{1308.4953}.

\bibitem[{\citenamefont{Tavakoli et~al.}(2014)\citenamefont{Tavakoli, Marto,
  and Dapor}}]{Tavakoli:2013rna}
\bibinfo{author}{\bibfnamefont{Y.}~\bibnamefont{Tavakoli}},
  \bibinfo{author}{\bibfnamefont{J.}~\bibnamefont{Marto}}, \bibnamefont{and}
  \bibinfo{author}{\bibfnamefont{A.}~\bibnamefont{Dapor}},
  \bibinfo{journal}{Int. J. Mod. Phys. D} \textbf{\bibinfo{volume}{23}},
  \bibinfo{pages}{1450061} (\bibinfo{year}{2014}), \eprint{1303.6157}.

\bibitem[{\citenamefont{Hayward}(2006)}]{Hayward:2005gi}
\bibinfo{author}{\bibfnamefont{S.~A.} \bibnamefont{Hayward}},
  \bibinfo{journal}{Phys. Rev. Lett.} \textbf{\bibinfo{volume}{96}},
  \bibinfo{pages}{031103} (\bibinfo{year}{2006}), \eprint{gr-qc/0506126}.

\bibitem[{\citenamefont{Goswami et~al.}(2006)\citenamefont{Goswami, Joshi, and
  Singh}}]{Goswami:2005fu}
\bibinfo{author}{\bibfnamefont{R.}~\bibnamefont{Goswami}},
  \bibinfo{author}{\bibfnamefont{P.~S.} \bibnamefont{Joshi}}, \bibnamefont{and}
  \bibinfo{author}{\bibfnamefont{P.}~\bibnamefont{Singh}},
  \bibinfo{journal}{Phys. Rev. Lett.} \textbf{\bibinfo{volume}{96}},
  \bibinfo{pages}{031302} (\bibinfo{year}{2006}), \eprint{gr-qc/0506129}.

\bibitem[{\citenamefont{Rovelli and Vidotto}(2014)}]{Rovelli:2014cta}
\bibinfo{author}{\bibfnamefont{C.}~\bibnamefont{Rovelli}} \bibnamefont{and}
  \bibinfo{author}{\bibfnamefont{F.}~\bibnamefont{Vidotto}},
  \bibinfo{journal}{Int. J. Mod. Phys. D} \textbf{\bibinfo{volume}{23}},
  \bibinfo{pages}{1442026} (\bibinfo{year}{2014}), \eprint{1401.6562}.

\bibitem[{\citenamefont{Papapetrou}(1985)}]{Dadhich:1985zx}
\bibinfo{author}{\bibfnamefont{A.}~\bibnamefont{Papapetrou}},
  \emph{\bibinfo{title}{{A random walk in Relativity and Cosmology. Essay in
  honnor of P. C. Vaidya and A. K. Raychaudhuri}}} (\bibinfo{year}{1985}).

\bibitem[{\citenamefont{Vaidya}(1999)}]{Vaidya:1999zz}
\bibinfo{author}{\bibfnamefont{P.~C.} \bibnamefont{Vaidya}},
  \bibinfo{journal}{Gen. Rel. Grav.} \textbf{\bibinfo{volume}{31}},
  \bibinfo{pages}{119} (\bibinfo{year}{1999}).

\bibitem[{\citenamefont{Wang and Wu}(1999{\natexlab{a}})}]{Wang:1998qx}
\bibinfo{author}{\bibfnamefont{A.}~\bibnamefont{Wang}} \bibnamefont{and}
  \bibinfo{author}{\bibfnamefont{Y.}~\bibnamefont{Wu}}, \bibinfo{journal}{Gen.
  Rel. Grav.} \textbf{\bibinfo{volume}{31}}, \bibinfo{pages}{107}
  (\bibinfo{year}{1999}{\natexlab{a}}), \eprint{gr-qc/9803038}.

\bibitem[{\citenamefont{Mkenyeleye et~al.}(2014)\citenamefont{Mkenyeleye,
  Goswami, and Maharaj}}]{Mkenyeleye:2014dwa}
\bibinfo{author}{\bibfnamefont{M.~D.} \bibnamefont{Mkenyeleye}},
  \bibinfo{author}{\bibfnamefont{R.}~\bibnamefont{Goswami}}, \bibnamefont{and}
  \bibinfo{author}{\bibfnamefont{S.~D.} \bibnamefont{Maharaj}},
  \bibinfo{journal}{Phys. Rev.} \textbf{\bibinfo{volume}{D90}},
  \bibinfo{pages}{064034} (\bibinfo{year}{2014}), \eprint{1407.4309}.

\bibitem[{\citenamefont{Harko and Cheng}(2000)}]{Harko:2000ni}
\bibinfo{author}{\bibfnamefont{T.}~\bibnamefont{Harko}} \bibnamefont{and}
  \bibinfo{author}{\bibfnamefont{K.}~\bibnamefont{Cheng}},
  \bibinfo{journal}{Phys. Lett. A} \textbf{\bibinfo{volume}{266}},
  \bibinfo{pages}{249} (\bibinfo{year}{2000}), \eprint{gr-qc/0104087}.

\bibitem[{\citenamefont{Ghosh and Dadhich}(2003)}]{Ghosh_2003}
\bibinfo{author}{\bibfnamefont{S.~G.} \bibnamefont{Ghosh}} \bibnamefont{and}
  \bibinfo{author}{\bibfnamefont{N.}~\bibnamefont{Dadhich}},
  \bibinfo{journal}{General Relativity and Gravitation}
  \textbf{\bibinfo{volume}{35}}, \bibinfo{pages}{359} (\bibinfo{year}{2003}).

\bibitem[{\citenamefont{Ghosh and Kothawala}(2008)}]{Ghosh:2008zza}
\bibinfo{author}{\bibfnamefont{S.}~\bibnamefont{Ghosh}} \bibnamefont{and}
  \bibinfo{author}{\bibfnamefont{D.}~\bibnamefont{Kothawala}},
  \bibinfo{journal}{Gen. Rel. Grav.} \textbf{\bibinfo{volume}{40}},
  \bibinfo{pages}{9} (\bibinfo{year}{2008}), \eprint{0801.4342}.

\bibitem[{\citenamefont{Husain}(1996)}]{Husain:1995bf}
\bibinfo{author}{\bibfnamefont{V.}~\bibnamefont{Husain}},
  \bibinfo{journal}{Phys. Rev.} \textbf{\bibinfo{volume}{D53}},
  \bibinfo{pages}{1759} (\bibinfo{year}{1996}), \eprint{gr-qc/9511011}.

\bibitem[{\citenamefont{Jhingan et~al.}(2001)\citenamefont{Jhingan, Dadhich,
  and Joshi}}]{Jhingan:2000xs}
\bibinfo{author}{\bibfnamefont{S.}~\bibnamefont{Jhingan}},
  \bibinfo{author}{\bibfnamefont{N.}~\bibnamefont{Dadhich}}, \bibnamefont{and}
  \bibinfo{author}{\bibfnamefont{P.~S.} \bibnamefont{Joshi}},
  \bibinfo{journal}{Phys. Rev.} \textbf{\bibinfo{volume}{D63}},
  \bibinfo{pages}{044010} (\bibinfo{year}{2001}), \eprint{gr-qc/0010111}.

\bibitem[{\citenamefont{Lake}(1991)}]{Lake:1991qrk}
\bibinfo{author}{\bibfnamefont{K.}~\bibnamefont{Lake}}, \bibinfo{journal}{Phys.
  Rev.} \textbf{\bibinfo{volume}{D43}}, \bibinfo{pages}{1416}
  (\bibinfo{year}{1991}).

\bibitem[{\citenamefont{Dwivedi and Joshi}(1989)}]{Dwivedi:1989pt}
\bibinfo{author}{\bibfnamefont{I.~H.} \bibnamefont{Dwivedi}} \bibnamefont{and}
  \bibinfo{author}{\bibfnamefont{P.~S.} \bibnamefont{Joshi}},
  \bibinfo{journal}{Class. Quant. Grav.} \textbf{\bibinfo{volume}{6}},
  \bibinfo{pages}{1599} (\bibinfo{year}{1989}).

\bibitem[{\citenamefont{Kuroda}(1984)}]{10.1143/PTP.72.63}
\bibinfo{author}{\bibfnamefont{Y.}~\bibnamefont{Kuroda}},
  \bibinfo{journal}{Progress of Theoretical Physics}
  \textbf{\bibinfo{volume}{72}}, \bibinfo{pages}{63} (\bibinfo{year}{1984}).

\bibitem[{\citenamefont{Sharif and Kausar}(2011)}]{Sharif:2010um}
\bibinfo{author}{\bibfnamefont{M.}~\bibnamefont{Sharif}} \bibnamefont{and}
  \bibinfo{author}{\bibfnamefont{H.}~\bibnamefont{Kausar}},
  \bibinfo{journal}{Astrophys. Space Sci.} \textbf{\bibinfo{volume}{331}},
  \bibinfo{pages}{281} (\bibinfo{year}{2011}), \eprint{1007.2852}.

\bibitem[{\citenamefont{Ziaie et~al.}(2011)\citenamefont{Ziaie, Atazadeh, and
  Rasouli}}]{Ziaie:2011dh}
\bibinfo{author}{\bibfnamefont{A.~H.} \bibnamefont{Ziaie}},
  \bibinfo{author}{\bibfnamefont{K.}~\bibnamefont{Atazadeh}}, \bibnamefont{and}
  \bibinfo{author}{\bibfnamefont{S.~M.~M.} \bibnamefont{Rasouli}},
  \bibinfo{journal}{Gen. Rel. Grav.} \textbf{\bibinfo{volume}{43}},
  \bibinfo{pages}{2943} (\bibinfo{year}{2011}), \eprint{1106.5638}.

\bibitem[{\citenamefont{Cembranos et~al.}(2012)\citenamefont{Cembranos,
  Cruz-Dombriz, and Núñez}}]{Cembranos_2012}
\bibinfo{author}{\bibfnamefont{J.}~\bibnamefont{Cembranos}},
  \bibinfo{author}{\bibfnamefont{A.~d.~l.} \bibnamefont{Cruz-Dombriz}},
  \bibnamefont{and} \bibinfo{author}{\bibfnamefont{B.~M.}
  \bibnamefont{Núñez}}, \bibinfo{journal}{Journal of Cosmology and
  Astroparticle Physics} \textbf{\bibinfo{volume}{2012}}, \bibinfo{pages}{021}
  (\bibinfo{year}{2012}).

\bibitem[{\citenamefont{Hwang and Yeom}(2010)}]{Hwang:2010aj}
\bibinfo{author}{\bibfnamefont{D.-i.} \bibnamefont{Hwang}} \bibnamefont{and}
  \bibinfo{author}{\bibfnamefont{D.-h.} \bibnamefont{Yeom}},
  \bibinfo{journal}{Class. Quant. Grav.} \textbf{\bibinfo{volume}{27}},
  \bibinfo{pages}{205002} (\bibinfo{year}{2010}), \eprint{1002.4246}.

\bibitem[{\citenamefont{Ziaie et~al.}(2010)\citenamefont{Ziaie, Atazadeh, and
  Tavakoli}}]{Ziaie:2010cz}
\bibinfo{author}{\bibfnamefont{A.~H.} \bibnamefont{Ziaie}},
  \bibinfo{author}{\bibfnamefont{K.}~\bibnamefont{Atazadeh}}, \bibnamefont{and}
  \bibinfo{author}{\bibfnamefont{Y.}~\bibnamefont{Tavakoli}},
  \bibinfo{journal}{Class. Quant. Grav.} \textbf{\bibinfo{volume}{27}},
  \bibinfo{pages}{075016} (\bibinfo{year}{2010}), \bibinfo{note}{[Erratum:
  Class. Quant. Grav.27,209801(2010)]}, \eprint{1003.1725}.

\bibitem[{\citenamefont{Bedjaoui et~al.}(2010)\citenamefont{Bedjaoui, LeFloch,
  Martin-Garcia, and Novak}}]{Bedjaoui:2010nh}
\bibinfo{author}{\bibfnamefont{N.}~\bibnamefont{Bedjaoui}},
  \bibinfo{author}{\bibfnamefont{P.~G.} \bibnamefont{LeFloch}},
  \bibinfo{author}{\bibfnamefont{J.~M.} \bibnamefont{Martin-Garcia}},
  \bibnamefont{and} \bibinfo{author}{\bibfnamefont{J.}~\bibnamefont{Novak}},
  \bibinfo{journal}{Class. Quant. Grav.} \textbf{\bibinfo{volume}{27}},
  \bibinfo{pages}{245010} (\bibinfo{year}{2010}), \eprint{1008.4238}.

\bibitem[{\citenamefont{Abbas and Tahir}(2017)}]{Abbas:2017tvh}
\bibinfo{author}{\bibfnamefont{G.}~\bibnamefont{Abbas}} \bibnamefont{and}
  \bibinfo{author}{\bibfnamefont{M.}~\bibnamefont{Tahir}},
  \bibinfo{journal}{Eur. Phys. J. C} \textbf{\bibinfo{volume}{77}},
  \bibinfo{pages}{537} (\bibinfo{year}{2017}), \eprint{1707.08472}.

\bibitem[{\citenamefont{Narita}(2009)}]{Narita:2009zza}
\bibinfo{author}{\bibfnamefont{M.}~\bibnamefont{Narita}}, \bibinfo{journal}{AIP
  Conf. Proc.} \textbf{\bibinfo{volume}{1122}}, \bibinfo{pages}{356}
  (\bibinfo{year}{2009}).

\bibitem[{\citenamefont{Ghosh and Jhingan}(2010)}]{Ghosh_2010}
\bibinfo{author}{\bibfnamefont{S.~G.} \bibnamefont{Ghosh}} \bibnamefont{and}
  \bibinfo{author}{\bibfnamefont{S.}~\bibnamefont{Jhingan}},
  \bibinfo{journal}{Physical Review D} \textbf{\bibinfo{volume}{82}}
  (\bibinfo{year}{2010}), ISSN \bibinfo{issn}{1550-2368}.

\bibitem[{\citenamefont{ZHOU et~al.}(2011)\citenamefont{ZHOU, YANG, ZOU, and
  YUE}}]{ZHOU_2011}
\bibinfo{author}{\bibfnamefont{K.}~\bibnamefont{ZHOU}},
  \bibinfo{author}{\bibfnamefont{Z.-Y.} \bibnamefont{YANG}},
  \bibinfo{author}{\bibfnamefont{D.-C.} \bibnamefont{ZOU}}, \bibnamefont{and}
  \bibinfo{author}{\bibfnamefont{R.-H.} \bibnamefont{YUE}},
  \bibinfo{journal}{Modern Physics Letters A} \textbf{\bibinfo{volume}{26}},
  \bibinfo{pages}{2135} (\bibinfo{year}{2011}).

\bibitem[{\citenamefont{Sharif and Abbas}(2013)}]{Sharif:2013era}
\bibinfo{author}{\bibfnamefont{M.}~\bibnamefont{Sharif}} \bibnamefont{and}
  \bibinfo{author}{\bibfnamefont{G.}~\bibnamefont{Abbas}},
  \bibinfo{journal}{Eur. Phys. J. Plus} \textbf{\bibinfo{volume}{128}},
  \bibinfo{pages}{102} (\bibinfo{year}{2013}), \eprint{1308.5675}.

\bibitem[{\citenamefont{Ohashi et~al.}(2011)\citenamefont{Ohashi, Shiromizu,
  and Jhingan}}]{Ohashi_2011}
\bibinfo{author}{\bibfnamefont{S.}~\bibnamefont{Ohashi}},
  \bibinfo{author}{\bibfnamefont{T.}~\bibnamefont{Shiromizu}},
  \bibnamefont{and} \bibinfo{author}{\bibfnamefont{S.}~\bibnamefont{Jhingan}},
  \bibinfo{journal}{Physical Review D} \textbf{\bibinfo{volume}{84}}
  (\bibinfo{year}{2011}), ISSN \bibinfo{issn}{1550-2368}.

\bibitem[{\citenamefont{Dadhich
  et~al.}(2013{\natexlab{b}})\citenamefont{Dadhich, Ghosh, and
  Jhingan}}]{Dadhich:2013bya}
\bibinfo{author}{\bibfnamefont{N.}~\bibnamefont{Dadhich}},
  \bibinfo{author}{\bibfnamefont{S.~G.} \bibnamefont{Ghosh}}, \bibnamefont{and}
  \bibinfo{author}{\bibfnamefont{S.}~\bibnamefont{Jhingan}},
  \bibinfo{journal}{Phys. Rev.} \textbf{\bibinfo{volume}{D88}},
  \bibinfo{pages}{084024} (\bibinfo{year}{2013}{\natexlab{b}}),
  \eprint{1308.4312}.

\bibitem[{\citenamefont{Zhou et~al.}(2011)\citenamefont{Zhou, Yang, Zou, and
  Yue}}]{Zhou:2011vz}
\bibinfo{author}{\bibfnamefont{K.}~\bibnamefont{Zhou}},
  \bibinfo{author}{\bibfnamefont{Z.-Y.} \bibnamefont{Yang}},
  \bibinfo{author}{\bibfnamefont{D.-C.} \bibnamefont{Zou}}, \bibnamefont{and}
  \bibinfo{author}{\bibfnamefont{R.-H.} \bibnamefont{Yue}},
  \bibinfo{journal}{Int. J. Mod. Phys. D} \textbf{\bibinfo{volume}{20}},
  \bibinfo{pages}{2317} (\bibinfo{year}{2011}), \eprint{1107.2730}.

\bibitem[{\citenamefont{{Abbas} and {Ahmed}}(2019)}]{G2019MPLA}
\bibinfo{author}{\bibfnamefont{G.}~\bibnamefont{{Abbas}}} \bibnamefont{and}
  \bibinfo{author}{\bibfnamefont{R.}~\bibnamefont{{Ahmed}}},
  \bibinfo{journal}{Modern Physics Letters A} \textbf{\bibinfo{volume}{34}},
  \bibinfo{eid}{1950153} (\bibinfo{year}{2019}).

\bibitem[{\citenamefont{Ahmed and Abbas}(2020)}]{Ahmed:2020xkh}
\bibinfo{author}{\bibfnamefont{R.}~\bibnamefont{Ahmed}} \bibnamefont{and}
  \bibinfo{author}{\bibfnamefont{G.}~\bibnamefont{Abbas}},
  \bibinfo{journal}{Mod. Phys. Lett. A} \textbf{\bibinfo{volume}{35}},
  \bibinfo{pages}{2050103} (\bibinfo{year}{2020}).

\bibitem[{\citenamefont{Ziaie et~al.}(2014{\natexlab{a}})\citenamefont{Ziaie,
  Moniz, Ranjbar, and Sepangi}}]{Ziaie_2014}
\bibinfo{author}{\bibfnamefont{A.~H.} \bibnamefont{Ziaie}},
  \bibinfo{author}{\bibfnamefont{P.~V.} \bibnamefont{Moniz}},
  \bibinfo{author}{\bibfnamefont{A.}~\bibnamefont{Ranjbar}}, \bibnamefont{and}
  \bibinfo{author}{\bibfnamefont{H.~R.} \bibnamefont{Sepangi}},
  \bibinfo{journal}{The European Physical Journal C}
  \textbf{\bibinfo{volume}{74}} (\bibinfo{year}{2014}{\natexlab{a}}).

\bibitem[{\citenamefont{Ziaie et~al.}(2014{\natexlab{b}})\citenamefont{Ziaie,
  Ranjbar, and Sepangi}}]{Ziaie_2014a}
\bibinfo{author}{\bibfnamefont{A.~H.} \bibnamefont{Ziaie}},
  \bibinfo{author}{\bibfnamefont{A.}~\bibnamefont{Ranjbar}}, \bibnamefont{and}
  \bibinfo{author}{\bibfnamefont{H.~R.} \bibnamefont{Sepangi}},
  \bibinfo{journal}{Classical and Quantum Gravity}
  \textbf{\bibinfo{volume}{32}}, \bibinfo{pages}{025010}
  (\bibinfo{year}{2014}{\natexlab{b}}).

\bibitem[{\citenamefont{Luz et~al.}(2018)\citenamefont{Luz, Mena, and
  Hadi~Ziaie}}]{Luz_2018}
\bibinfo{author}{\bibfnamefont{P.}~\bibnamefont{Luz}},
  \bibinfo{author}{\bibfnamefont{F.~C.} \bibnamefont{Mena}}, \bibnamefont{and}
  \bibinfo{author}{\bibfnamefont{A.}~\bibnamefont{Hadi~Ziaie}},
  \bibinfo{journal}{Classical and Quantum Gravity}
  \textbf{\bibinfo{volume}{36}}, \bibinfo{pages}{015003}
  (\bibinfo{year}{2018}).

\bibitem[{\citenamefont{Frolov}(2004)}]{PhysRevD.70.104023}
\bibinfo{author}{\bibfnamefont{A.~V.} \bibnamefont{Frolov}},
  \bibinfo{journal}{Phys. Rev. D} \textbf{\bibinfo{volume}{70}},
  \bibinfo{pages}{104023} (\bibinfo{year}{2004}).

\bibitem[{\citenamefont{Gutperle and Kraus}(2004)}]{Gutperle_2004}
\bibinfo{author}{\bibfnamefont{M.}~\bibnamefont{Gutperle}} \bibnamefont{and}
  \bibinfo{author}{\bibfnamefont{P.}~\bibnamefont{Kraus}},
  \bibinfo{journal}{Journal of High Energy Physics}
  \textbf{\bibinfo{volume}{2004}}, \bibinfo{pages}{024} (\bibinfo{year}{2004}).

\bibitem[{\citenamefont{Horowitz}(2005)}]{Horowitz:2003yv}
\bibinfo{author}{\bibfnamefont{G.~T.} \bibnamefont{Horowitz}},
  \bibinfo{journal}{Phys. Scripta T} \textbf{\bibinfo{volume}{117}},
  \bibinfo{pages}{86} (\bibinfo{year}{2005}), \eprint{hep-th/0312123}.

\bibitem[{\citenamefont{Bonanno et~al.}(2018)\citenamefont{Bonanno, Koch, and
  Platania}}]{Bonanno:2017zen}
\bibinfo{author}{\bibfnamefont{A.}~\bibnamefont{Bonanno}},
  \bibinfo{author}{\bibfnamefont{B.}~\bibnamefont{Koch}}, \bibnamefont{and}
  \bibinfo{author}{\bibfnamefont{A.}~\bibnamefont{Platania}},
  \bibinfo{journal}{Found. Phys.} \textbf{\bibinfo{volume}{48}},
  \bibinfo{pages}{1393} (\bibinfo{year}{2018}), \eprint{1710.10845}.

\bibitem[{\citenamefont{Platania}(2018)}]{platania2018asymptotically}
\bibinfo{author}{\bibfnamefont{A.}~\bibnamefont{Platania}},
  \emph{\bibinfo{title}{Asymptotically Safe Gravity: From Spacetime Foliation
  to Cosmology}}, Springer Theses (\bibinfo{publisher}{Springer International
  Publishing}, \bibinfo{year}{2018}).

\bibitem[{\citenamefont{Reuter and Saueressig}(2019)}]{reuter2019quantum}
\bibinfo{author}{\bibfnamefont{M.}~\bibnamefont{Reuter}} \bibnamefont{and}
  \bibinfo{author}{\bibfnamefont{F.}~\bibnamefont{Saueressig}},
  \emph{\bibinfo{title}{Quantum Gravity and the Functional Renormalization
  Group: The Road towards Asymptotic Safety}}, Cambridge Monographs on
  Mathematical Physics (\bibinfo{publisher}{Cambridge University Press},
  \bibinfo{year}{2019}).

\bibitem[{\citenamefont{Bonanno et~al.}(2017)\citenamefont{Bonanno, Koch, and
  Platania}}]{Bonanno:2016dyv}
\bibinfo{author}{\bibfnamefont{A.}~\bibnamefont{Bonanno}},
  \bibinfo{author}{\bibfnamefont{B.}~\bibnamefont{Koch}}, \bibnamefont{and}
  \bibinfo{author}{\bibfnamefont{A.}~\bibnamefont{Platania}},
  \bibinfo{journal}{Class. Quant. Grav.} \textbf{\bibinfo{volume}{34}},
  \bibinfo{pages}{095012} (\bibinfo{year}{2017}), \eprint{1610.05299}.

\bibitem[{\citenamefont{Arkani-Hamed et~al.}(2007)\citenamefont{Arkani-Hamed,
  Motl, Nicolis, and Vafa}}]{ArkaniHamed:2006dz}
\bibinfo{author}{\bibfnamefont{N.}~\bibnamefont{Arkani-Hamed}},
  \bibinfo{author}{\bibfnamefont{L.}~\bibnamefont{Motl}},
  \bibinfo{author}{\bibfnamefont{A.}~\bibnamefont{Nicolis}}, \bibnamefont{and}
  \bibinfo{author}{\bibfnamefont{C.}~\bibnamefont{Vafa}},
  \bibinfo{journal}{JHEP} \textbf{\bibinfo{volume}{06}}, \bibinfo{pages}{060}
  (\bibinfo{year}{2007}), \eprint{hep-th/0601001}.

\bibitem[{\citenamefont{Horowitz and Santos}(2019)}]{Horowitz:2019eum}
\bibinfo{author}{\bibfnamefont{G.~T.} \bibnamefont{Horowitz}} \bibnamefont{and}
  \bibinfo{author}{\bibfnamefont{J.~E.} \bibnamefont{Santos}},
  \bibinfo{journal}{JHEP} \textbf{\bibinfo{volume}{06}}, \bibinfo{pages}{122}
  (\bibinfo{year}{2019}), \eprint{1901.11096}.

\bibitem[{\citenamefont{Horowitz et~al.}(2016)\citenamefont{Horowitz, Santos,
  and Way}}]{Horowitz:2016ezu}
\bibinfo{author}{\bibfnamefont{G.~T.} \bibnamefont{Horowitz}},
  \bibinfo{author}{\bibfnamefont{J.~E.} \bibnamefont{Santos}},
  \bibnamefont{and} \bibinfo{author}{\bibfnamefont{B.}~\bibnamefont{Way}},
  \bibinfo{journal}{Class. Quant. Grav.} \textbf{\bibinfo{volume}{33}},
  \bibinfo{pages}{195007} (\bibinfo{year}{2016}), \eprint{1604.06465}.

\bibitem[{\citenamefont{Crisford and Santos}(2017)}]{Crisford:2017zpi}
\bibinfo{author}{\bibfnamefont{T.}~\bibnamefont{Crisford}} \bibnamefont{and}
  \bibinfo{author}{\bibfnamefont{J.~E.} \bibnamefont{Santos}},
  \bibinfo{journal}{Phys. Rev. Lett.} \textbf{\bibinfo{volume}{118}},
  \bibinfo{pages}{181101} (\bibinfo{year}{2017}), \eprint{1702.05490}.

\bibitem[{\citenamefont{Ong}(2020)}]{Ong:2020xwv}
\bibinfo{author}{\bibfnamefont{Y.~C.} \bibnamefont{Ong}}
  (\bibinfo{year}{2020}), \eprint{2005.07032}.

\bibitem[{\citenamefont{Lobo}(2015)}]{Lobo:2014ara}
\bibinfo{author}{\bibfnamefont{F.~S.~N.} \bibnamefont{Lobo}},
  \bibinfo{journal}{J. Phys. Conf. Ser.} \textbf{\bibinfo{volume}{600}},
  \bibinfo{pages}{012006} (\bibinfo{year}{2015}), \eprint{1412.0867}.

\bibitem[{\citenamefont{Faraoni and Capozziello}(2011)}]{Capozziello:2010zz}
\bibinfo{author}{\bibfnamefont{V.}~\bibnamefont{Faraoni}} \bibnamefont{and}
  \bibinfo{author}{\bibfnamefont{S.}~\bibnamefont{Capozziello}},
  \emph{\bibinfo{title}{{Beyond Einstein Gravity}}}, vol. \bibinfo{volume}{170}
  (\bibinfo{publisher}{Springer}, \bibinfo{address}{Dordrecht},
  \bibinfo{year}{2011}).

\bibitem[{\citenamefont{Parker}(1971)}]{Parker:1971pt}
\bibinfo{author}{\bibfnamefont{L.}~\bibnamefont{Parker}},
  \bibinfo{journal}{Phys. Rev.} \textbf{\bibinfo{volume}{D3}},
  \bibinfo{pages}{346} (\bibinfo{year}{1971}), \bibinfo{note}{[Erratum: Phys.
  Rev.D3,2546(1971)]}.

\bibitem[{\citenamefont{Rastall}(1972)}]{Rastall:1973nw}
\bibinfo{author}{\bibfnamefont{P.}~\bibnamefont{Rastall}},
  \bibinfo{journal}{Phys. Rev.} \textbf{\bibinfo{volume}{D6}},
  \bibinfo{pages}{3357} (\bibinfo{year}{1972}).

\bibitem[{\citenamefont{Harko and Lobo}(2014)}]{Harko:2014gwa}
\bibinfo{author}{\bibfnamefont{T.}~\bibnamefont{Harko}} \bibnamefont{and}
  \bibinfo{author}{\bibfnamefont{F.~S.~N.} \bibnamefont{Lobo}},
  \bibinfo{journal}{Galaxies} \textbf{\bibinfo{volume}{2}},
  \bibinfo{pages}{410} (\bibinfo{year}{2014}), \eprint{1407.2013}.

\bibitem[{\citenamefont{Bertolami et~al.}(2007)\citenamefont{Bertolami,
  Boehmer, Harko, and Lobo}}]{Bertolami:2007gv}
\bibinfo{author}{\bibfnamefont{O.}~\bibnamefont{Bertolami}},
  \bibinfo{author}{\bibfnamefont{C.~G.} \bibnamefont{Boehmer}},
  \bibinfo{author}{\bibfnamefont{T.}~\bibnamefont{Harko}}, \bibnamefont{and}
  \bibinfo{author}{\bibfnamefont{F.~S.~N.} \bibnamefont{Lobo}},
  \bibinfo{journal}{Phys. Rev.} \textbf{\bibinfo{volume}{D75}},
  \bibinfo{pages}{104016} (\bibinfo{year}{2007}), \eprint{0704.1733}.

\bibitem[{\citenamefont{Heydarzade and Darabi}(2017)}]{Heydarzade_2017}
\bibinfo{author}{\bibfnamefont{Y.}~\bibnamefont{Heydarzade}} \bibnamefont{and}
  \bibinfo{author}{\bibfnamefont{F.}~\bibnamefont{Darabi}},
  \bibinfo{journal}{Physics Letters B} \textbf{\bibinfo{volume}{771}},
  \bibinfo{pages}{365} (\bibinfo{year}{2017}), ISSN \bibinfo{issn}{0370-2693}.

\bibitem[{\citenamefont{Kiselev}(2003)}]{Kiselev_2003}
\bibinfo{author}{\bibfnamefont{V.~V.} \bibnamefont{Kiselev}},
  \bibinfo{journal}{Classical and Quantum Gravity}
  \textbf{\bibinfo{volume}{20}}, \bibinfo{pages}{1187} (\bibinfo{year}{2003}),
  ISSN \bibinfo{issn}{0264-9381}.

\bibitem[{\citenamefont{Fabris et~al.}(2012)\citenamefont{Fabris, Piattella,
  Rodrigues, Batista, and Daouda}}]{FABRIS_2012}
\bibinfo{author}{\bibfnamefont{J.~C.} \bibnamefont{Fabris}},
  \bibinfo{author}{\bibfnamefont{O.~F.} \bibnamefont{Piattella}},
  \bibinfo{author}{\bibfnamefont{D.~C.} \bibnamefont{Rodrigues}},
  \bibinfo{author}{\bibfnamefont{C.~E.~M.} \bibnamefont{Batista}},
  \bibnamefont{and} \bibinfo{author}{\bibfnamefont{M.~H.}
  \bibnamefont{Daouda}}, \bibinfo{journal}{International Journal of Modern
  Physics: Conference Series} \textbf{\bibinfo{volume}{18}},
  \bibinfo{pages}{67} (\bibinfo{year}{2012}), ISSN \bibinfo{issn}{2010-1945}.

\bibitem[{\citenamefont{Moradpour
  et~al.}(2017{\natexlab{a}})\citenamefont{Moradpour, Heydarzade, Darabi, and
  Salako}}]{Moradpour:2017shy}
\bibinfo{author}{\bibfnamefont{H.}~\bibnamefont{Moradpour}},
  \bibinfo{author}{\bibfnamefont{Y.}~\bibnamefont{Heydarzade}},
  \bibinfo{author}{\bibfnamefont{F.}~\bibnamefont{Darabi}}, \bibnamefont{and}
  \bibinfo{author}{\bibfnamefont{I.~G.} \bibnamefont{Salako}},
  \bibinfo{journal}{Eur. Phys. J.} \textbf{\bibinfo{volume}{C77}},
  \bibinfo{pages}{259} (\bibinfo{year}{2017}{\natexlab{a}}),
  \eprint{1704.02458}.

\bibitem[{\citenamefont{Al-Rawaf and
  Taha}(1996{\natexlab{a}})}]{AlRawaf:1995rs}
\bibinfo{author}{\bibfnamefont{A.~S.} \bibnamefont{Al-Rawaf}} \bibnamefont{and}
  \bibinfo{author}{\bibfnamefont{M.~O.} \bibnamefont{Taha}},
  \bibinfo{journal}{Phys. Lett.} \textbf{\bibinfo{volume}{B366}},
  \bibinfo{pages}{69} (\bibinfo{year}{1996}{\natexlab{a}}).

\bibitem[{\citenamefont{Abdel-Rahman}(1997)}]{Rahman1997}
\bibinfo{author}{\bibfnamefont{A.}~\bibnamefont{Abdel-Rahman}},
  \bibinfo{journal}{Gen. Rel. Grav.} \textbf{\bibinfo{volume}{29}},
  \bibinfo{pages}{1329} (\bibinfo{year}{1997}).

\bibitem[{\citenamefont{Arbab}(2003)}]{Arbab_2003}
\bibinfo{author}{\bibfnamefont{A.~I.} \bibnamefont{Arbab}},
  \bibinfo{journal}{Journal of Cosmology and Astroparticle Physics}
  \textbf{\bibinfo{volume}{2003}}, \bibinfo{pages}{008} (\bibinfo{year}{2003}),
  ISSN \bibinfo{issn}{1475-7516}.

\bibitem[{\citenamefont{Batista et~al.}(2010)\citenamefont{Batista, Fabris, and
  Daouda}}]{Batista:2010nq}
\bibinfo{author}{\bibfnamefont{C.~E.~M.} \bibnamefont{Batista}},
  \bibinfo{author}{\bibfnamefont{J.~C.} \bibnamefont{Fabris}},
  \bibnamefont{and} \bibinfo{author}{\bibfnamefont{M.~H.}
  \bibnamefont{Daouda}}, \bibinfo{journal}{Nuovo Cim.}
  \textbf{\bibinfo{volume}{B125}}, \bibinfo{pages}{957} (\bibinfo{year}{2010}),
  \eprint{1004.4603}.

\bibitem[{\citenamefont{Batista et~al.}(2013)\citenamefont{Batista, Fabris,
  Piattella, and Velasquez-Toribio}}]{Batista:2012hv}
\bibinfo{author}{\bibfnamefont{C.~E.~M.} \bibnamefont{Batista}},
  \bibinfo{author}{\bibfnamefont{J.~C.} \bibnamefont{Fabris}},
  \bibinfo{author}{\bibfnamefont{O.~F.} \bibnamefont{Piattella}},
  \bibnamefont{and} \bibinfo{author}{\bibfnamefont{A.~M.}
  \bibnamefont{Velasquez-Toribio}}, \bibinfo{journal}{Eur. Phys. J.}
  \textbf{\bibinfo{volume}{C73}}, \bibinfo{pages}{2425} (\bibinfo{year}{2013}),
  \eprint{1208.6327}.

\bibitem[{\citenamefont{Al-Rawaf and
  Taha}(1996{\natexlab{b}})}]{AlRawaf:1994pn}
\bibinfo{author}{\bibfnamefont{A.~S.} \bibnamefont{Al-Rawaf}} \bibnamefont{and}
  \bibinfo{author}{\bibfnamefont{M.~O.} \bibnamefont{Taha}},
  \bibinfo{journal}{Gen. Rel. Grav.} \textbf{\bibinfo{volume}{28}},
  \bibinfo{pages}{935} (\bibinfo{year}{1996}{\natexlab{b}}).

\bibitem[{\citenamefont{Batista et~al.}(2012)\citenamefont{Batista, Daouda,
  Fabris, Piattella, and Rodrigues}}]{Batista:2011nu}
\bibinfo{author}{\bibfnamefont{C.~E.~M.} \bibnamefont{Batista}},
  \bibinfo{author}{\bibfnamefont{M.~H.} \bibnamefont{Daouda}},
  \bibinfo{author}{\bibfnamefont{J.~C.} \bibnamefont{Fabris}},
  \bibinfo{author}{\bibfnamefont{O.~F.} \bibnamefont{Piattella}},
  \bibnamefont{and} \bibinfo{author}{\bibfnamefont{D.~C.}
  \bibnamefont{Rodrigues}}, \bibinfo{journal}{Phys. Rev.}
  \textbf{\bibinfo{volume}{D85}}, \bibinfo{pages}{084008}
  (\bibinfo{year}{2012}), \eprint{1112.4141}.

\bibitem[{\citenamefont{Majernik}(2003)}]{Majernik:2002gd}
\bibinfo{author}{\bibfnamefont{V.}~\bibnamefont{Majernik}},
  \bibinfo{journal}{Gen. Rel. Grav.} \textbf{\bibinfo{volume}{35}},
  \bibinfo{pages}{1007} (\bibinfo{year}{2003}), \eprint{gr-qc/0201019}.

\bibitem[{\citenamefont{Abdel-Rahman}(2001)}]{AbdelRahman:2001pb}
\bibinfo{author}{\bibfnamefont{A.~M.~M.} \bibnamefont{Abdel-Rahman}},
  \bibinfo{journal}{Astrophys. Space Sci.} \textbf{\bibinfo{volume}{278}},
  \bibinfo{pages}{383} (\bibinfo{year}{2001}), \bibinfo{note}{[Astrophys. Space
  Sci.278,385(2001)]}.

\bibitem[{\citenamefont{Visser}(2018)}]{Visser:2017gpz}
\bibinfo{author}{\bibfnamefont{M.}~\bibnamefont{Visser}},
  \bibinfo{journal}{Phys. Lett.} \textbf{\bibinfo{volume}{B782}},
  \bibinfo{pages}{83} (\bibinfo{year}{2018}), \eprint{1711.11500}.

\bibitem[{\citenamefont{Gratus et~al.}(2012)\citenamefont{Gratus, Obukhov, and
  Tucker}}]{Gratus_2012}
\bibinfo{author}{\bibfnamefont{J.}~\bibnamefont{Gratus}},
  \bibinfo{author}{\bibfnamefont{Y.~N.} \bibnamefont{Obukhov}},
  \bibnamefont{and} \bibinfo{author}{\bibfnamefont{R.~W.}
  \bibnamefont{Tucker}}, \bibinfo{journal}{Annals of Physics}
  \textbf{\bibinfo{volume}{327}}, \bibinfo{pages}{2560} (\bibinfo{year}{2012}).

\bibitem[{\citenamefont{Darabi et~al.}(2018)\citenamefont{Darabi, Moradpour,
  Licata, Heydarzade, and Corda}}]{Darabi_2018}
\bibinfo{author}{\bibfnamefont{F.}~\bibnamefont{Darabi}},
  \bibinfo{author}{\bibfnamefont{H.}~\bibnamefont{Moradpour}},
  \bibinfo{author}{\bibfnamefont{I.}~\bibnamefont{Licata}},
  \bibinfo{author}{\bibfnamefont{Y.}~\bibnamefont{Heydarzade}},
  \bibnamefont{and} \bibinfo{author}{\bibfnamefont{C.}~\bibnamefont{Corda}},
  \bibinfo{journal}{The European Physical Journal C}
  \textbf{\bibinfo{volume}{78}} (\bibinfo{year}{2018}), ISSN
  \bibinfo{issn}{1434-6052}.

\bibitem[{\citenamefont{Moradpour
  et~al.}(2017{\natexlab{b}})\citenamefont{Moradpour, Bonilla, Abreu, and
  Neto}}]{PhysRevD.96.123504}
\bibinfo{author}{\bibfnamefont{H.}~\bibnamefont{Moradpour}},
  \bibinfo{author}{\bibfnamefont{A.}~\bibnamefont{Bonilla}},
  \bibinfo{author}{\bibfnamefont{E.~M.~C.} \bibnamefont{Abreu}},
  \bibnamefont{and} \bibinfo{author}{\bibfnamefont{J.~A.} \bibnamefont{Neto}},
  \bibinfo{journal}{Phys. Rev. D} \textbf{\bibinfo{volume}{96}},
  \bibinfo{pages}{123504} (\bibinfo{year}{2017}{\natexlab{b}}).

\bibitem[{\citenamefont{Smalley}(1983)}]{Smalley_1983}
\bibinfo{author}{\bibfnamefont{L.~L.} \bibnamefont{Smalley}},
  \bibinfo{journal}{Journal of Physics A: Mathematical and General}
  \textbf{\bibinfo{volume}{16}}, \bibinfo{pages}{2179} (\bibinfo{year}{1983}).

\bibitem[{\citenamefont{Hansraj et~al.}(2019)\citenamefont{Hansraj, Banerjee,
  and Channuie}}]{Hansraj_2019}
\bibinfo{author}{\bibfnamefont{S.}~\bibnamefont{Hansraj}},
  \bibinfo{author}{\bibfnamefont{A.}~\bibnamefont{Banerjee}}, \bibnamefont{and}
  \bibinfo{author}{\bibfnamefont{P.}~\bibnamefont{Channuie}},
  \bibinfo{journal}{Annals of Physics} \textbf{\bibinfo{volume}{400}},
  \bibinfo{pages}{320} (\bibinfo{year}{2019}), ISSN \bibinfo{issn}{0003-4916}.

\bibitem[{\citenamefont{Tolman}(1939)}]{PhysRev.55.364}
\bibinfo{author}{\bibfnamefont{R.~C.} \bibnamefont{Tolman}},
  \bibinfo{journal}{Phys. Rev.} \textbf{\bibinfo{volume}{55}},
  \bibinfo{pages}{364} (\bibinfo{year}{1939}).

\bibitem[{\citenamefont{Oliveira et~al.}(2015)\citenamefont{Oliveira, Velten,
  Fabris, and Casarini}}]{PhysRevD.92.044020}
\bibinfo{author}{\bibfnamefont{A.~M.} \bibnamefont{Oliveira}},
  \bibinfo{author}{\bibfnamefont{H.~E.~S.} \bibnamefont{Velten}},
  \bibinfo{author}{\bibfnamefont{J.~C.} \bibnamefont{Fabris}},
  \bibnamefont{and} \bibinfo{author}{\bibfnamefont{L.}~\bibnamefont{Casarini}},
  \bibinfo{journal}{Phys. Rev. D} \textbf{\bibinfo{volume}{92}},
  \bibinfo{pages}{044020} (\bibinfo{year}{2015}).

\bibitem[{\citenamefont{Abbas and Shahzad}(2018)}]{Abbas_2018}
\bibinfo{author}{\bibfnamefont{G.}~\bibnamefont{Abbas}} \bibnamefont{and}
  \bibinfo{author}{\bibfnamefont{M.~R.} \bibnamefont{Shahzad}},
  \bibinfo{journal}{The European Physical Journal A}
  \textbf{\bibinfo{volume}{54}} (\bibinfo{year}{2018}), ISSN
  \bibinfo{issn}{1434-601X}.

\bibitem[{ABB(2020)}]{ABBAS20201}
\bibinfo{journal}{Chinese Journal of Physics} \textbf{\bibinfo{volume}{63}},
  \bibinfo{pages}{1 } (\bibinfo{year}{2020}), ISSN \bibinfo{issn}{0577-9073}.

\bibitem[{\citenamefont{Ziaie et~al.}(2019)\citenamefont{Ziaie, Moradpour, and
  Ghaffari}}]{Ziaie_2019}
\bibinfo{author}{\bibfnamefont{A.~H.} \bibnamefont{Ziaie}},
  \bibinfo{author}{\bibfnamefont{H.}~\bibnamefont{Moradpour}},
  \bibnamefont{and} \bibinfo{author}{\bibfnamefont{S.}~\bibnamefont{Ghaffari}},
  \bibinfo{journal}{Physics Letters B} \textbf{\bibinfo{volume}{793}},
  \bibinfo{pages}{276} (\bibinfo{year}{2019}).

\bibitem[{\citenamefont{Capone et~al.}(2010)\citenamefont{Capone, Cardone, and
  Ruggiero}}]{Capone_2010}
\bibinfo{author}{\bibfnamefont{M.}~\bibnamefont{Capone}},
  \bibinfo{author}{\bibfnamefont{V.~F.} \bibnamefont{Cardone}},
  \bibnamefont{and} \bibinfo{author}{\bibfnamefont{M.~L.}
  \bibnamefont{Ruggiero}}, \bibinfo{journal}{Journal of Physics: Conference
  Series} \textbf{\bibinfo{volume}{222}}, \bibinfo{pages}{012012}
  (\bibinfo{year}{2010}).

\bibitem[{\citenamefont{Moradpour and Salako}(2016)}]{Moradpour:2016fur}
\bibinfo{author}{\bibfnamefont{H.}~\bibnamefont{Moradpour}} \bibnamefont{and}
  \bibinfo{author}{\bibfnamefont{I.~G.} \bibnamefont{Salako}},
  \bibinfo{journal}{Adv. High Energy Phys.} \textbf{\bibinfo{volume}{2016}},
  \bibinfo{pages}{3492796} (\bibinfo{year}{2016}), \eprint{1606.06589}.

\bibitem[{\citenamefont{Bonnor and Vaidya}(1970)}]{Bonnor:1970zz}
\bibinfo{author}{\bibfnamefont{W.}~\bibnamefont{Bonnor}} \bibnamefont{and}
  \bibinfo{author}{\bibfnamefont{P.}~\bibnamefont{Vaidya}},
  \bibinfo{journal}{Gen. Rel. Grav.} \textbf{\bibinfo{volume}{1}},
  \bibinfo{pages}{127} (\bibinfo{year}{1970}).

\bibitem[{\citenamefont{Joshi and Malafarina}(2011)}]{JOSHI_2011}
\bibinfo{author}{\bibfnamefont{P.~S.} \bibnamefont{Joshi}} \bibnamefont{and}
  \bibinfo{author}{\bibfnamefont{D.}~\bibnamefont{Malafarina}},
  \bibinfo{journal}{International Journal of Modern Physics D}
  \textbf{\bibinfo{volume}{20}}, \bibinfo{pages}{2641–2729}
  (\bibinfo{year}{2011}).

\bibitem[{\citenamefont{Poisson}(2004)}]{poisson_2004}
\bibinfo{author}{\bibfnamefont{E.}~\bibnamefont{Poisson}},
  \emph{\bibinfo{title}{A Relativist's Toolkit: The Mathematics of Black-Hole
  Mechanics}} (\bibinfo{publisher}{Cambridge University Press},
  \bibinfo{year}{2004}).

\bibitem[{\citenamefont{Wang and Wu}(1999{\natexlab{b}})}]{Wang_1999}
\bibinfo{author}{\bibfnamefont{A.}~\bibnamefont{Wang}} \bibnamefont{and}
  \bibinfo{author}{\bibfnamefont{Y.}~\bibnamefont{Wu}},
  \bibinfo{journal}{General Relativity and Gravitation}
  \textbf{\bibinfo{volume}{31}}, \bibinfo{pages}{107}
  (\bibinfo{year}{1999}{\natexlab{b}}).

\bibitem[{\citenamefont{Nolan et~al.}(2002)\citenamefont{Nolan, Mena, and
  Gonçalves}}]{Nolan_2002}
\bibinfo{author}{\bibfnamefont{B.~C.} \bibnamefont{Nolan}},
  \bibinfo{author}{\bibfnamefont{F.~C.} \bibnamefont{Mena}}, \bibnamefont{and}
  \bibinfo{author}{\bibfnamefont{S.~M.} \bibnamefont{Gonçalves}},
  \bibinfo{journal}{Physics Letters A} \textbf{\bibinfo{volume}{294}},
  \bibinfo{pages}{122–125} (\bibinfo{year}{2002}), ISSN
  \bibinfo{issn}{0375-9601}.

\bibitem[{\citenamefont{Maeda}(2006)}]{Maeda_2006}
\bibinfo{author}{\bibfnamefont{H.}~\bibnamefont{Maeda}},
  \bibinfo{journal}{Classical and Quantum Gravity}
  \textbf{\bibinfo{volume}{23}}, \bibinfo{pages}{2155} (\bibinfo{year}{2006}),
  ISSN \bibinfo{issn}{1361-6382}.

\bibitem[{\citenamefont{York}(1984)}]{York:1984}
\bibinfo{author}{\bibfnamefont{J.~W.~J.} \bibnamefont{York}},
  \emph{\bibinfo{title}{{in Quantum theory of gravity : essays in honor of the
  60th birthday of Bryce S. DeWitt, Page 135 }}} (\bibinfo{publisher}{Bristol
  [Avon] : A. Hilger}, \bibinfo{year}{1984}).

\bibitem[{\citenamefont{Faraoni}(2018)}]{Faraoni:2018xwo}
\bibinfo{author}{\bibfnamefont{V.}~\bibnamefont{Faraoni}},
  \bibinfo{journal}{Universe} \textbf{\bibinfo{volume}{4}},
  \bibinfo{pages}{109} (\bibinfo{year}{2018}), \eprint{1810.04667}.

\bibitem[{\citenamefont{Tipler}(1977)}]{Tipler:1977zza}
\bibinfo{author}{\bibfnamefont{F.~J.} \bibnamefont{Tipler}},
  \bibinfo{journal}{Phys. Lett.} \textbf{\bibinfo{volume}{A64}},
  \bibinfo{pages}{8} (\bibinfo{year}{1977}).

\bibitem[{\citenamefont{Ori}(1991)}]{Ori:1991zz}
\bibinfo{author}{\bibfnamefont{A.}~\bibnamefont{Ori}}, \bibinfo{journal}{Phys.
  Rev. Lett.} \textbf{\bibinfo{volume}{67}}, \bibinfo{pages}{789}
  (\bibinfo{year}{1991}).

\bibitem[{\citenamefont{Clarke and Kr\'olak}(1985)}]{Clarke:1985}
\bibinfo{author}{\bibfnamefont{C.~J.~S.} \bibnamefont{Clarke}}
  \bibnamefont{and} \bibinfo{author}{\bibfnamefont{A.}~\bibnamefont{Kr\'olak}},
  \bibinfo{journal}{J. Geom. Phys.} \textbf{\bibinfo{volume}{2}},
  \bibinfo{pages}{127} (\bibinfo{year}{1985}).

\bibitem[{\citenamefont{Barriola and Vilenkin}(1989)}]{Barriola:1989hx}
\bibinfo{author}{\bibfnamefont{M.}~\bibnamefont{Barriola}} \bibnamefont{and}
  \bibinfo{author}{\bibfnamefont{A.}~\bibnamefont{Vilenkin}},
  \bibinfo{journal}{Phys. Rev. Lett.} \textbf{\bibinfo{volume}{63}},
  \bibinfo{pages}{341} (\bibinfo{year}{1989}).

\bibitem[{\citenamefont{Witten}(1984)}]{Witten:1984rs}
\bibinfo{author}{\bibfnamefont{E.}~\bibnamefont{Witten}},
  \bibinfo{journal}{Phys. Rev. D} \textbf{\bibinfo{volume}{30}},
  \bibinfo{pages}{272} (\bibinfo{year}{1984}).

\bibitem[{\citenamefont{Weinberg}(1996)}]{weinberg_1996}
\bibinfo{author}{\bibfnamefont{S.}~\bibnamefont{Weinberg}},
  \emph{\bibinfo{title}{The Quantum Theory of Fields}},
  vol.~\bibinfo{volume}{2} (\bibinfo{publisher}{Cambridge University Press},
  \bibinfo{year}{1996}).

\end{thebibliography}

\end{document}